\documentclass[%
 reprint,
 amsmath,amssymb,
 aps,
 physrev,
floatfix,
]{revtex4-2}
\usepackage{mathtools}
\usepackage[utf8]{inputenc}
\usepackage{textalpha}
\usepackage{newunicodechar}
\newunicodechar{₀}{\textsubscript{0}}
\newunicodechar{₁}{\textsubscript{1}}
\newunicodechar{₂}{\textsubscript{2}}
\newunicodechar{₃}{\textsubscript{3}}
\newunicodechar{₄}{\textsubscript{4}}
\newunicodechar{₅}{\textsubscript{5}}
\newunicodechar{₆}{\textsubscript{6}}
\newunicodechar{₇}{\textsubscript{7}}
\newunicodechar{₈}{\textsubscript{8}}
\newunicodechar{₉}{\textsubscript{9}}
\newunicodechar{⁰}{\textsuperscript{0}}
\newunicodechar{⁴}{\textsuperscript{4}}
\newunicodechar{⁵}{\textsuperscript{5}}
\newunicodechar{⁶}{\textsuperscript{6}}
\newunicodechar{⁷}{\textsuperscript{7}}
\newunicodechar{⁸}{\textsuperscript{8}}
\newunicodechar{⁹}{\textsuperscript{9}}
\newunicodechar{α}{\ensuremath{\alpha}}
\usepackage{graphicx}
\usepackage{dcolumn}
\usepackage{bm}
\usepackage{glossaries}
\usepackage{glossaries-extra}
\usepackage{xr}
\usepackage[hidelinks]{hyperref}
\usepackage{orcidlink}
\usepackage{hyperref}  
\usepackage[capitalize]{cleveref}

\crefname{equation}{Eq.}{Eqs.}
\Crefname{equation}{Equation}{Equations}

\crefname{figure}{Fig.}{Figs.}
\Crefname{figure}{Figure}{Figures}

\crefname{table}{Table}{Tables}
\Crefname{table}{Table}{Tables}
\usepackage{needspace}
\newcommand{\arxivtighten}{\looseness=-1{}}
\newcommand{\arxivsqueeze}{\vspace{-4pt}}
\usepackage{xcolor}
\usepackage{tikz}
\usetikzlibrary{external, calc}
\tikzset{external/mode=graphics if exists}
\usepackage{contour}

\newacronym{dft}{DFT}{density-functional theory}
\newacronym{dftu}{DFT+\emph{U}}{Hubbard-corrected DFT}
\newacronym[first=DFT+\emph{U}+\emph{V}]{dftuv}{DFT+\emph{U}+\emph{V}}{to include inter-site corrections}
\newacronym{sie}{SIE}{self-interaction error}
\newacronym[plural=TMOs,firstplural=transition metal oxides (TMOs)]{tmo}{TMO}{transition metal oxide}
\newcommand{\trace}[1]{\text{Tr}\left[#1\right]}
\newcommand{\bra}[1]{\langle #1 |}
\newcommand{\ket}[1]{| #1 \rangle}
\newcommand{\braket}[2]{\langle #1 | #2 \rangle}
\newcommand{\braopket}[3]{\langle #1 | #2 | #3 \rangle}
\newcommand{\maintitle}{DFT+$U$+$V$ is equivalent to DFT+$U$ with density-dependent hybridized projectors}

\usepackage[switch,mathlines]{lineno}
\newif\ifrefereelines\refereelinestrue
\usepackage{etoolbox}
\AtBeginEnvironment{tikzpicture}{\nolinenumbers}
\BeforeBeginEnvironment{align}{\linenomath}
\AfterEndEnvironment{align}{\endlinenomath}
\refereelinesfalse


\crefname{section}{Supp.~Mat.}{Supp.~Mat.}
\Crefname{section}{Supp.~Mat.}{Supp.~Mat.}

\begin{document}
\ifrefereelines\linenumbers\fi

\title{\maintitle}

\author{Edward Linscott\orcidlink{0000-0002-4967-9873}}
 \email{edward.linscott@psi.ch}
\affiliation{%
 PSI Center for Scientific Computing, Theory and Data, Paul Scherrer Institute, 5232 Villigen PSI, Switzerland}%
\affiliation{%
 National Centre for Computational Design and Discovery of Novel Materials (MARVEL), Paul Scherrer Institute, 5232 Villigen PSI, Switzerland}%
\author{Alberto Carta\orcidlink{0000-0003-0705-0281}}
\affiliation{%
 PSI Center for Scientific Computing, Theory and Data, Paul Scherrer Institute, 5232 Villigen PSI, Switzerland}%
\affiliation{%
 National Centre for Computational Design and Discovery of Novel Materials (MARVEL), Paul Scherrer Institute, 5232 Villigen PSI, Switzerland}%
\author{Nicola Marzari\orcidlink{0000-0002-9764-0199}}%
\affiliation{%
 PSI Center for Scientific Computing, Theory and Data, Paul Scherrer Institute, 5232 Villigen PSI, Switzerland}%
\affiliation{%
 National Centre for Computational Design and Discovery of Novel Materials (MARVEL), Paul Scherrer Institute, 5232 Villigen PSI, Switzerland}%
\affiliation{%
Theory and Simulation of Materials (THEOS), École Polytechnique Fédérale de Lausanne, 1015 Lausanne, Switzerland}%
\affiliation{%
Theory of Condensed Matter, Cavendish Laboratory, University of Cambridge, Cambridge CB3 0US, United Kingdom
\vspace{2ex}}

\date{\today}

\begin{abstract}%
  \glsunsetall
  \noindent Hubbard-corrected density-functional theory (\gls{dftu}) is a popular tool for first-principles modeling of materials with localized $d$ or $f$ electrons, but its on-site corrections tend to over-localize charge and break covalent bonds. Inter-site $+V$ corrections were introduced to counter this and are now widely used, but a formal justification has been lacking. Here we show that --- to first order in $V/U$ --- inter-site corrections are exactly equivalent to on-site \gls{dftu} evaluated on a density-matrix-dependent redefinition of the Hubbard projectors, hybridized with those of neighboring sites, providing insight into the explicit mechanism by which $V$ affects covalency. If the projectors are held frozen, as is common practice, the equivalence partially breaks down. Beyond reinterpreting the formalism, these results sharpen the questions of how $V$ should be computed and what the Hubbard subspaces fundamentally are.
\end{abstract}

\maketitle

\glsresetall
\noindent \Acrfull{dft} is a cornerstone of first-principles modeling across physics, chemistry, and materials science~\cite{Hohenberg1964a,Kohn1965a,Jones2015a,Marzari2021,VanNoorden2025}, but it fails for so-called ``strongly correlated'' materials --- loosely, those whose ground-state wavefunction has no single dominant electronic configuration --- for instance, wrongly predicting many \glspl{tmo} to be metallic~\cite{Terakura1984,Anisimov1991a}.
\Gls{dftu} was initially developed to address this failure, inspired by the success of the Hubbard model (a comparatively simple model Hamiltonian that captures the essential physics of the Mott transition \cite{Hubbard1963a}).
\linelabel{hunds_j:start}The \gls{dftu} energy functional --- here in its rotationally invariant simplified form, using fully-localized-limit double-counting and with the Hund's coupling $J$ absorbed into an effective Hubbard parameter~\cite{Liechtenstein1995,Dudarev1998a,Czyzyk1994} --- is given by\linelabel{hunds_j:end}
%
\begin{equation}
E_{\text{DFT}+U} = E_{\text{DFT}} + \sum_{I\sigma} \frac{U^I}{2} \trace{\mathbf{n}^{II\sigma} (1 - \mathbf{n}^{II\sigma})},
\label{eq:e_dftu}
\end{equation}
where the occupancy matrix
\begin{equation}
n^{II\sigma}_{mm'} = \braopket{\phi^{I\sigma}_m}{\hat \rho^\sigma}{\phi^{I\sigma}_{m'}} = \sum_\nu f_\nu^\sigma \braket{\phi^{I\sigma}_m}{\psi_\nu^\sigma} \braket{\psi_\nu^\sigma}{\phi^{I\sigma}_{m'}}
\end{equation}
is the projection of the spin-dependent Kohn-Sham density-matrix operator $\hat{\rho}^\sigma = \sum_\nu f_\nu^\sigma \ket{\psi_\nu^\sigma}\bra{\psi_\nu^\sigma}$ (with $\nu$ being a compound index for both crystal momentum and band index) onto a local subspace (indexed $I$) as defined by a set of orbitals $\{\phi^{I\sigma}_m\}$ \cite{Anisimov1993a,Anisimov1997a,Pickett1998a}. Typically, these orbitals are atom-centered, fixed, localized, orthonormal (as in the ``ortho-atomic'' scheme employed in this work), and correspond to the $3d$ or $4f$ subshell of a transition metal/lanthanide~\footnote{They are also usually spin-independent, but we have assigned them a spin index $\sigma$ for the sake of generality.}. 
Meanwhile, $U^I$ is the Hubbard parameter for site $I$, and defines the strength of the Hubbard correction that will be applied to this particular subspace.
\linelabel{gks:start}Because the correction depends on the Kohn-Sham orbitals via the projected density matrix, \gls{dftu} is an orbital-dependent, generalized Kohn-Sham theory.\linelabel{gks:end}

The derivation of \gls{dftu} from the Hubbard model rests on several severe approximations --- chief among them, that all electronic interactions are neglected except for local Hartree-Fock terms within each Hubbard site.
\linelabel{degrade:start}Consequently, \gls{dftu} does not model genuinely strongly correlated chemistry --- indeed, in some cases it degrades predictions relative to uncorrected \gls{dft}~\cite{Campo2010a,Kulik2011a,Jain2011} --- but for the most part it has been applied with remarkable success to many different classes of materials (see Refs.~\cite{Scherlis2007a,Hsu2011,Bajdich2013}, among many others).\linelabel{degrade:end}

\arxivtighten%
\linelabel{point4:start}\gls{dftu} can be placed on more rigorous and transparent footing by framing it as a correction not of strong correlations, but of self-interaction \cite{Cococcioni2005a,Kulik2006,Kulik2015b}. This alternative interpretation rests on an exact condition: for the exact functional, the total energy is piecewise linear in the total number of electrons in the system, whereas approximate functionals exhibit erroneous curvature~\cite{Perdew1982,Yang2000}. This is known as many-electron \gls{sie} or delocalization error \cite{Mori-Sanchez2006,Mori-Sanchez2008}. The key observation of Refs.~\cite{Cococcioni2005a,Kulik2006} is that the Hubbard correction of \cref{eq:e_dftu} directly reduces the curvature of the total energy with respect to the occupations of each Hubbard subspace (more specifically, the eigenvalues $\lambda^{I\sigma}_i$ of the on-site occupation matrices $\mathbf{n}^{II\sigma}$). With a judicious choice of $U^I$, the \gls{dftu} energy term can therefore remove this local curvature entirely. This is an ansatz, not an exact condition: correcting local curvature may, but need not, remedy the \emph{global} curvature with respect to total particle number~\cite{Zhao2016a}.\linelabel{point4:end}
\linelabel{linear_bath:start}It can be partly justified by noting that if the rest of the system interacts with the Hubbard site only through an energy that is linear in the site occupation --- as it does if the on-site electronic interactions are so strong that it is a good approximation to neglect all other electron-electron interactions, or if the rest of the system acts through a constant potential --- the exact total energy is then piecewise linear in the subspace occupation itself, so any local curvature is genuinely spurious and the ansatz becomes exact~\cite{Burgess2024b}.\linelabel{linear_bath:end}

Framing \gls{dftu} as a correction to \gls{sie} not only helps to explain its success, but it also provides us with a class of recipes to calculate the Hubbard parameters $\{U^I\}$: compute the local curvature directly from linear response \gls{dft}, and then set $\{U^I\}$ to match the observed error (see \cref{sec:linear_response_details}~\cite{SuppMat}). Since \gls{dftu} predictions depend strongly on the value of $U$, an unbiased first-principles procedure is a prerequisite for the theory to be predictive.

However, even with a recipe for computing Hubbard parameters, the \gls{dftu} method is not without its shortcomings --- not least that the choice of the Hubbard projectors themselves remains largely arbitrary~\cite{Kirchner-Hall2021}, a point to which we will return. In addition, it has been observed that \gls{dftu} can over-localize charge and artificially break covalent bonds~\cite{Kulik2011a,Kulik2011b}. This arises because the Hubbard corrections shift the energy levels of the projector orbitals, disrupting any hybridization (\emph{i.e.}\ covalency) that might be occurring between these orbitals and those of neighboring atoms (because effective hybridization between orbitals requires them to have similar energies)~\cite{Linscott2018,Moore2024}. This is not a corner case: away from the most ionic compounds, metal--ligand hybridization is the rule rather than the exception, and a purely on-site correction cannot see it. In an effort to address this, \gls{dftu} has been extended
to \gls{dftuv}:
\begin{equation}
  E_{\mathrm{DFT}+U+V} =
  E_{\mathrm{DFT}+U}
  - {\sum_{IJ\sigma}} \frac{V^{IJ}}{2} \trace{\mathbf{n}^{IJ\sigma} \mathbf{n}^{JI\sigma}},
  \label{eq:e_dftuv}
\end{equation}
where the strength of these inter-site corrections is defined by the $V^{IJ}$ parameters~\cite{Campo2010a,Kulik2011a}. Typically, $V^{IJ}$ is chosen to be non-zero only between neighboring sites. Like the \gls{dftu} functional, it follows from the Hubbard model under a single-Slater-determinant approximation \cite{Dudarev1998a}, now also retaining Hartree-Fock interactions between adjacent sites.

$V$ parameters are typically calculated from the off-diagonal components of the same response matrices constructed to determine $U$. However, the formal justification for doing so is unclear: while the diagonal response measures the local curvature of the total energy with respect to the occupation of the Hubbard subspace, $\frac{d^2E}{d\lambda_i^2}$, which the $+U$ framework treats as spurious and can therefore be framed as a correction to \gls{sie}, there is no equivalent reason why the off-diagonal $\frac{d^2E}{d\lambda_i d\lambda_j}$ should also be zero.
Put another way, $+U$ has a clear target --- removing the spurious on-site curvature $\frac{d^2E}{d\lambda_i^2}$ --- whereas it is far less clear what error of approximate \gls{dft} the $+V$ correction is meant to remedy.
\linelabel{mismatch_letter:start}There is also a mismatch of a more basic kind: the response matrices measure curvature with respect to the site-diagonal occupations $n^{II}$ and $n^{JJ}$, whereas $E_V$ introduces curvature with respect to the inter-site blocks, $\partial^2 E_V / \partial n^{IJ\sigma}_{mm'} \partial n^{JI\sigma}_{m'm} = -V^{IJ}$ --- so the quantity computed and the quantity corrected are not the same object~\cite{Bastonero2025}.\linelabel{mismatch_letter:end}
Nevertheless, \gls{dftuv} has proven remarkably successful in practice, improving the description of band gaps, oxidation states, cathode voltages, and defect chemistry~\cite{Cococcioni2019,Ricca2020,Tancogne-Dejean2020,Mahajan2022,Timrov2022a,Yang2022b,Binci2023a,Kam2025}. Its advantage over \gls{dftu} can occasionally be marginal~\cite{Yang2026,Haddadi2024,Lee2020}, but its quantitative accuracy can be excellent: computed intercalation voltages of olivine-type Li-ion cathodes fall within a few percent of experiment, outperforming \gls{dft}, \gls{dftu}, and HSE06 alike~\cite{Timrov2022a}. This raises the question: what is the physical meaning of the $V$ corrections, and why do they typically improve over \gls{dftu}?

\begin{figure*}[t]
\tikzsetnextfilename{fig-isosurfaces}
\begin{tikzpicture}[x=0.11224489795\textwidth, y=0.11224489795\textwidth, every node/.style={inner sep=0pt}]
  \foreach \x in {0,1,2}
  {%
    \pgfmathsetmacro{\temp}{3 - \x}
    \pgfmathtruncatemacro{\y}{\temp}

    \node[anchor=south west] at (1, \x) {\includegraphics[width=0.10204081632\textwidth]{figures/isosurfaces/diagonalizing/NiO.dn_0000\y.png}};
    \node[anchor=south west] at (2, \x) {\includegraphics[width=0.10204081632\textwidth]{figures/isosurfaces/hybridized/NiO.dn_0000\y.png}};
  }
  \node[anchor=south west] at (0, 2) {\includegraphics[width=0.10204081632\textwidth]{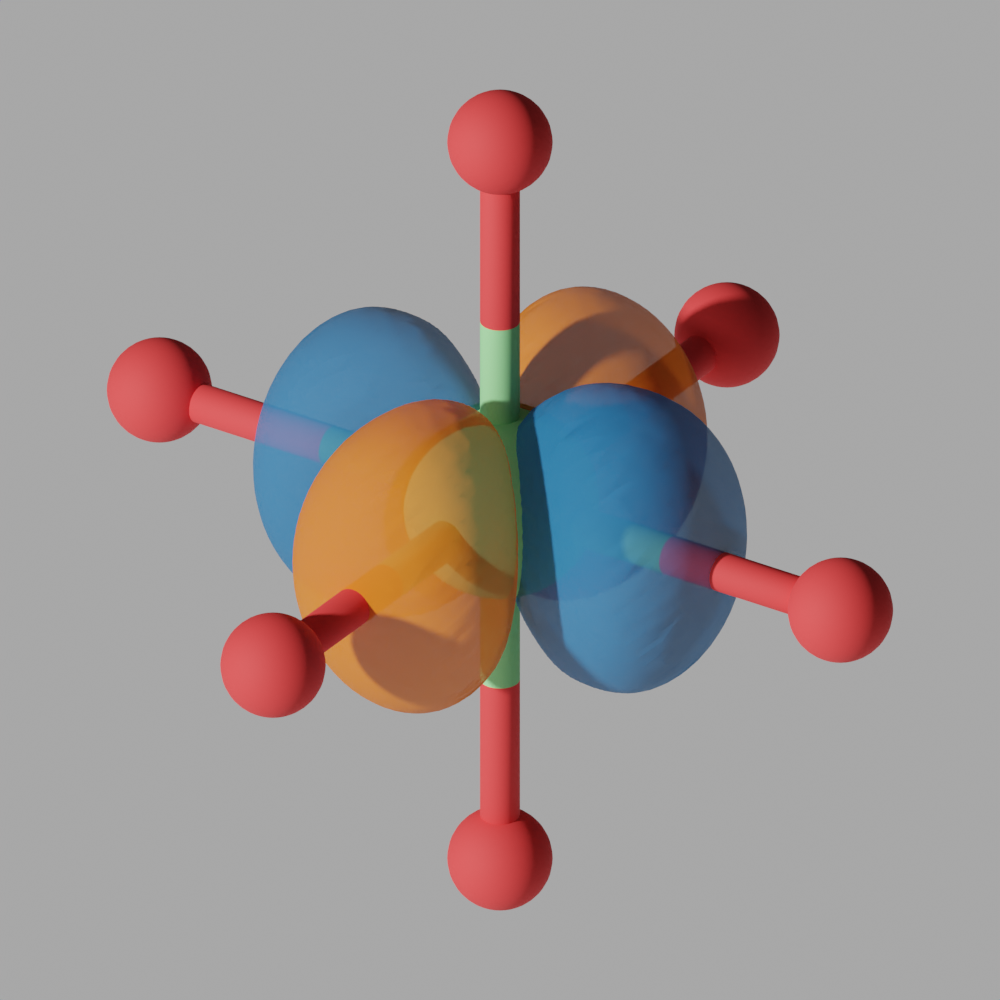}};
  \node[anchor=south west] at (0, 1) {\includegraphics[width=0.10204081632\textwidth]{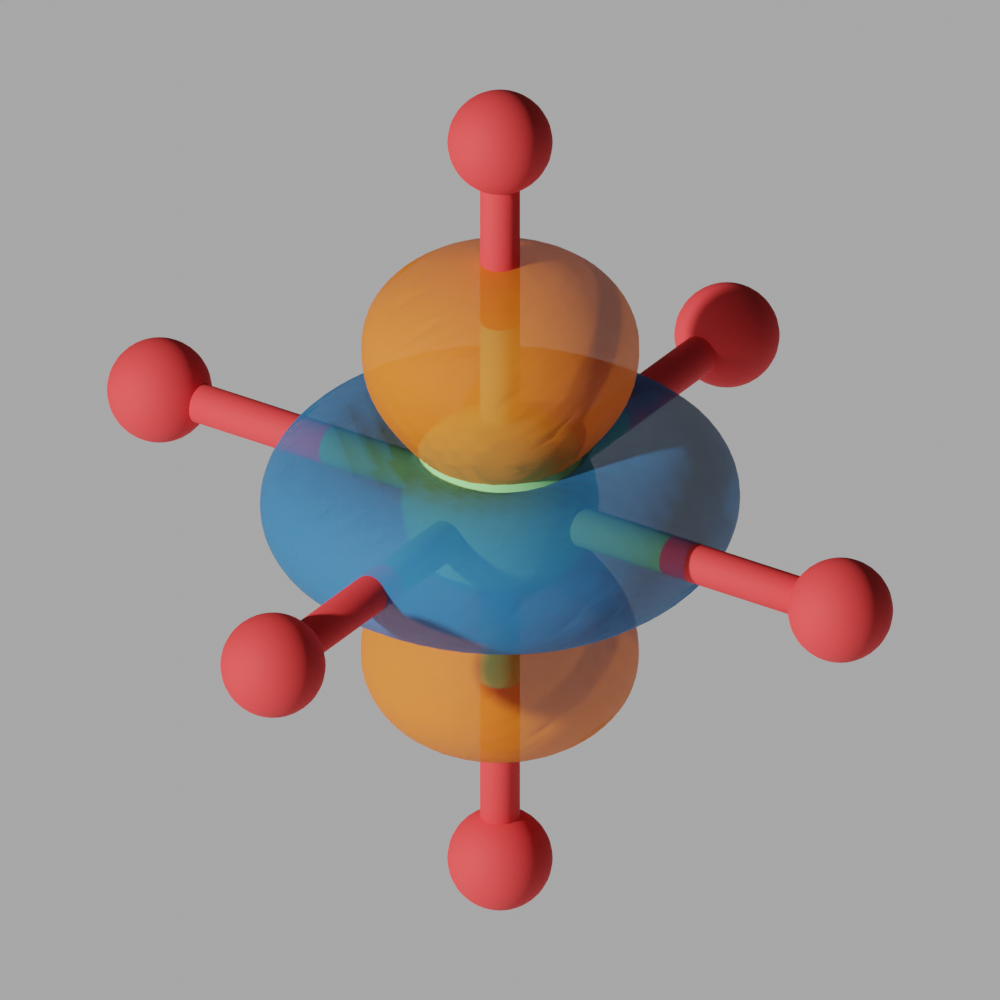}};
  \node[anchor=south west] at (0, 0) {\includegraphics[width=0.10204081632\textwidth]{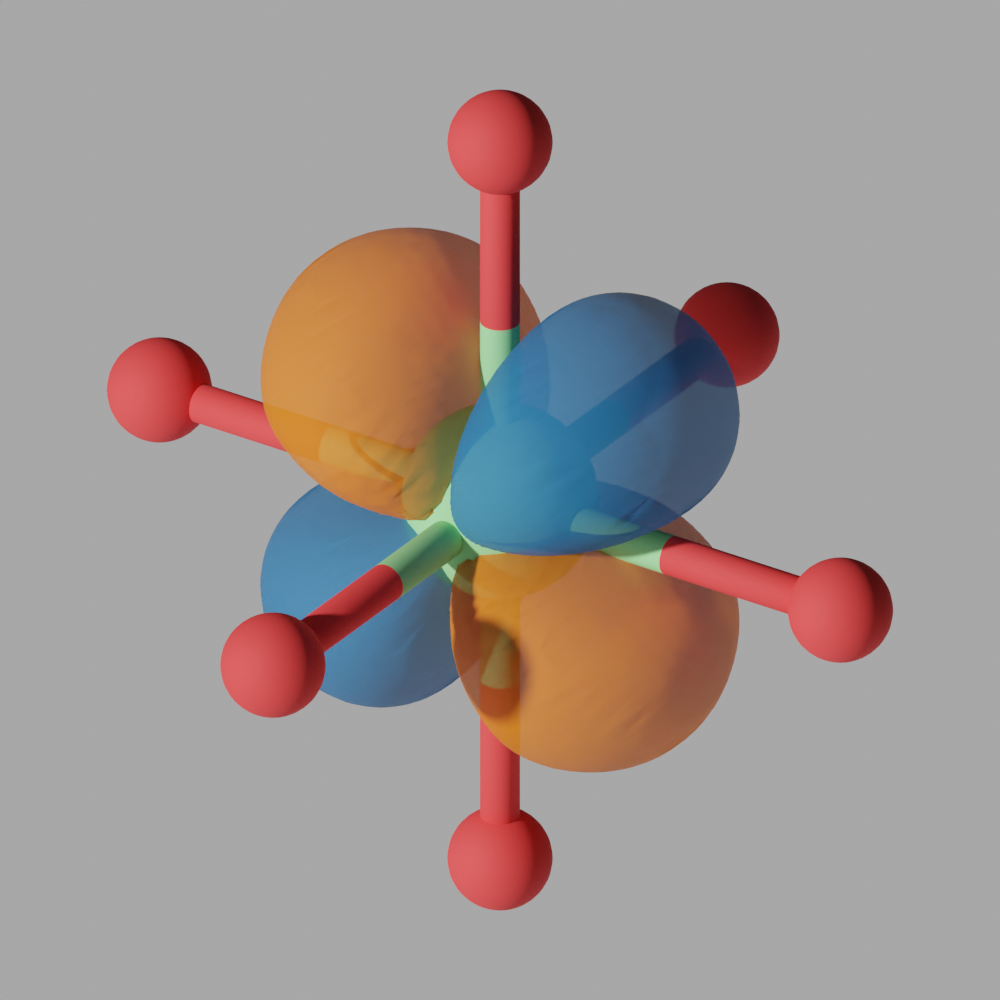}};

  \node[anchor=south west] at (3, 0) {\includegraphics[width=0.32653061224\textwidth]{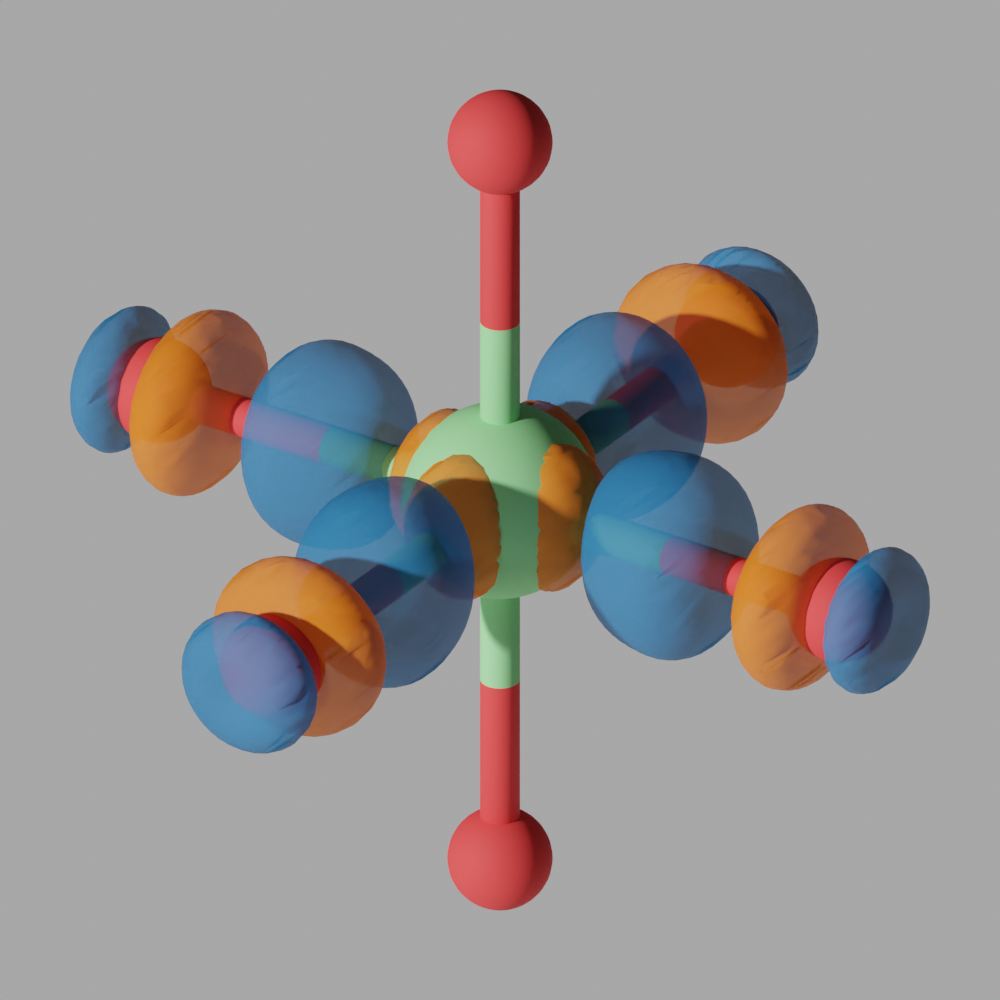}};
  \node[anchor=south west] at (6, 0) {\includegraphics[width=0.32653061224\textwidth]{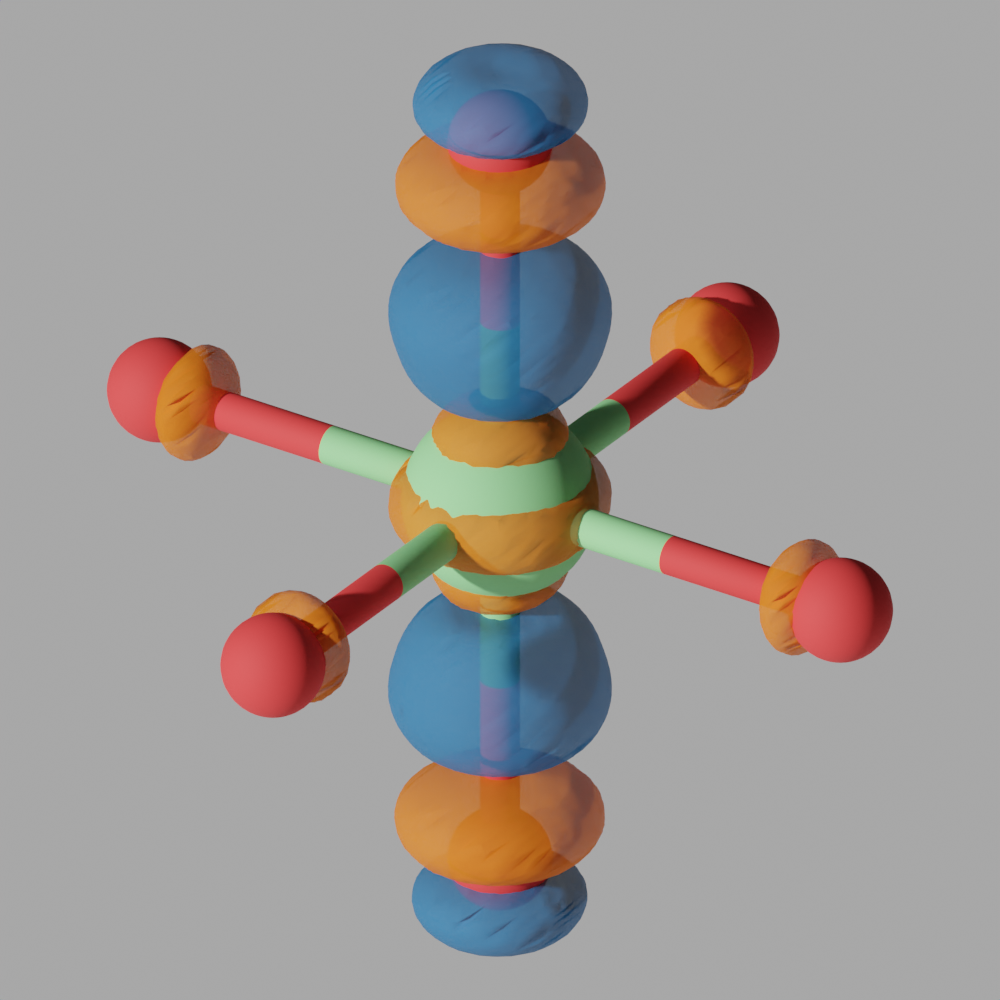}};

  \node at (0.4545, 3.05) {$\phi^{I\sigma}_m(\mathbf{r})$};
  \node at (1.4545, 3.05) {$\varphi^{I\sigma}_i(\mathbf{r})$};
  \node at (2.4545, 3.05) {$\tilde{\varphi}^{I\sigma}_i(\mathbf{r})$};
  \node at (4.4545, 3.05) {$|\tilde{\varphi}^{I\sigma}_1(\mathbf{r})|^2 - |\varphi^{I\sigma}_1(\mathbf{r})|^2$};
  \node at (7.4545, 3.05) {$|\tilde{\varphi}^{I\sigma}_2(\mathbf{r})|^2 - |\varphi^{I\sigma}_2(\mathbf{r})|^2$};

  \node[anchor=west] at (0.05, 2.75) {\large \textsf{a}\vphantom{abcdefghijklmno}};
  \node[anchor=west] at (0.05, 1.75) {\large \textsf{b}\vphantom{abcdefghijklmno}};
  \node[anchor=west] at (0.05, 0.75) {\large \textsf{c}\vphantom{abcdefghijklmno}};
  \node[anchor=west] at (1.05, 2.75) {\large \textsf{d}\vphantom{abcdefghijklmno}};
  \node[anchor=west] at (1.05, 1.75) {\large \textsf{e}\vphantom{abcdefghijklmno}};
  \node[anchor=west] at (1.05, 0.75) {\large \textsf{f}\vphantom{abcdefghijklmno}};
  \node[anchor=west] at (2.05, 2.75) {\large \textsf{g}\vphantom{abcdefghijklmno}};
  \node[anchor=west] at (2.05, 1.75) {\large \textsf{h}\vphantom{abcdefghijklmno}};
  \node[anchor=west] at (2.05, 0.75) {\large \textsf{i}\vphantom{abcdefghijklmno}};
  \node[anchor=west] at (3.05, 2.75) {\large \textsf{j}\vphantom{abcdefghijklmno}};
  \node[anchor=west] at (6.05, 2.75) {\large \textsf{k}\vphantom{abcdefghijklmno}};

\end{tikzpicture}
\caption{The original $\phi^{I\sigma}_m(\mathbf{r})$, block-diagonalizing $\varphi^{I\sigma}_i(\mathbf{r})$, and doubly hybridized Hubbard projectors $\tilde{\varphi}^{I\sigma}_i(\mathbf{r}) = e^{2\hat S}\varphi^{I\sigma}_i(\mathbf{r})$ of nickel in antiferromagnetic NiO for the minority spin channel. For simplicity, only one nickel atom (green) and its nearest-neighbor oxygen atoms (red) are shown. The left column (a--c) shows three of the five original Hubbard projectors, defined as orthogonalized pseudo-atomic orbitals oriented with the Cartesian axes. (The crystal structure happens to align with the Cartesian directions, but in general the orientation of these orbitals is arbitrary.) The second column (d--f) shows the block-diagonalizing projectors $\ket{\varphi^{I\sigma}_i}$, \emph{i.e.}\ the eigenvectors of $\mathbf{n}^{II}$; unlike the original projectors, these are oriented with respect to the local chemical environment, with the octahedral crystal field and the exchange field of the antiferromagnetic ordering producing a $d_{x^2-y^2}$ orbital (d), a $d_{z^2}$ orbital (e), and a combination of the degenerate $t_{2g}$ orbitals (f). The third column (g--i) shows the doubly hybridized projectors; the first two weakly hybridize with the oxygen $2p$ orbitals. Panels j and k show the density difference between the first two hybridized and block-diagonalizing orbitals (orange regions positive, blue negative), revealing the increased antibonding character with the chemically relevant oxygens.}
\label{fig:all_orbitals}
\end{figure*}

This Letter demonstrates that to first order in $V/U$, \gls{dftuv} is exactly on-site \gls{dftu} evaluated on a redefinition of the Hubbard projectors. To prove this, we first rewrite the \gls{dftu} energy in the basis $\{\ket{\varphi_i^{I\sigma}}\}$ of eigenvectors of the on-site occupation matrix $\mathbf{n}^{II\sigma}$, with eigenvalues $\lambda_i^{I\sigma}$. These ``block-diagonalizing projectors'' are the natural orbitals of each site --- oriented by the local chemical environment rather than by the arbitrary orientation of the original projectors --- and their occupations $\lambda_i^{I\sigma}$ are precisely the quantities whose spurious curvature the $+U$ correction removes. The block-diagonalizing projectors of antiferromagnetic NiO are shown in Figs.~\ref{fig:all_orbitals}d--\ref{fig:all_orbitals}f alongside the original projectors (panels a--c). We then hybridize each site with its $V$-coupled neighbors by rotating
%
$
  \ket{\varphi_i^{I\sigma}} \longrightarrow e^{\hat T}\ket{\varphi_i^{I\sigma}},
$
where $\hat T$ is an as-yet-unknown anti-Hermitian generator that induces a weak hybridization between neighboring sites, built from the Hubbard parameters and the inter-site occupations $\Lambda^{IJ\sigma}_{ij} = \braopket{\varphi_i^{I\sigma}}{\hat\rho^\sigma}{\varphi_j^{J\sigma}}$ (the off-diagonal density-matrix elements that quantify the covalency between sites $I$ and $J$). In \hyperref[app:proof]{the Appendix} we prove that there is a specific choice $\hat{T} = \hat{S}$ for which the \gls{dftu} energy of the hybridized projectors reproduces the \gls{dftuv} energy,
\begin{equation}
  E_U[\hat S] = E_U + E_V + \mathcal{O}(V^2/U),
  \label{eq:energy_identity}
\end{equation}
\arxivtighten%
where $E_U[\hat S]$ denotes the \gls{dftu} correction energy evaluated using the ``hybridized'' projectors $\{e^{\hat S}\ket{\varphi_i^{I\sigma}}\}$, and $E_U + E_V$ is the energy of the \gls{dftuv} correction applied to the original projectors. This generator $\hat{S}$ is given explicitly by
\arxivsqueeze%
\begin{align}
  \bra{\varphi_i^{I\sigma}}\hat{S}\ket{\varphi_j^{J\sigma}}
  =& \frac{V^{IJ} \Lambda^{IJ\sigma}_{ij}}{U^I(1 - 2 \lambda_i^{I\sigma}) - U^J(1 - 2\lambda_j^{J\sigma})}.
  \label{eq:explicit_s}
\end{align}
(For bulk systems, $\hat{S}$ takes the $\mathbf{k}$-dependent form given in \cref{sec:periodic}.)
\linelabel{vu_note:start}The expansion is controlled by the norm of the generator $\|\hat{S}\|$, which, for typical magnitudes $V$ and $U$ of the inter-site couplings and on-site potentials, is of order $V/U$ (see \cref{sec:s_norm} for details). For reference, in bulk NiO, $\|\hat{S}\| \approx 0.08$.\linelabel{vu_note:end}

\Cref{eq:energy_identity} can be differentiated, upon which we find that, to first order, the \gls{dftu} potential applied to these hybridized projectors reproduces the \gls{dftuv} potential. This is the central result of this work: in the limit $V \ll U$, the inter-site $+V$ correction is exactly a redefinition of the Hubbard projectors. Since $\hat{S}$ is built from the occupations themselves, this redefinition is intrinsically \emph{density-matrix-dependent}. This also identifies the error that $+V$ remedies: to first order it imposes the very same occupation-curvature penalty as $+U$ --- a correction to \gls{sie} --- but measured on the hybridized projectors rather than the bare ones.

\arxivtighten%
To understand how this rotation affects the Hubbard projectors, consider one common \gls{dftuv} setup, in which inter-site corrections couple a $3d$ or $4f$ site to the $2p$ orbitals of ligands that are not themselves Hubbard-corrected. For this case, the rotation acts on the projectors as
\begin{equation}
  \ket{\varphi_i^{I\sigma}} \longrightarrow \ket{\varphi_i^{I\sigma}} - \sum_{J} \sum_j \frac{\Lambda^{JI\sigma}_{ji}}{1 - 2\lambda_i^{I\sigma}} \frac{V^{IJ}}{U^I} \ket{\varphi_j^{J\sigma}}.
  \label{eq:delta_phi_uj0}
\end{equation}
\Cref{eq:delta_phi_uj0} shows that the new hybridized projectors remain centered on site $I$ but acquire weight from the orbitals of its $V$-coupled neighbors (\emph{e.g.}\ the $2p$ orbitals of oxygen in a \glsfmtlong{tmo}). The admixture of $\ket{\varphi_j^{J\sigma}}$ grows with $V^{IJ}/U^I$ and with the off-diagonal occupancy $\Lambda^{IJ\sigma}_{ij}$, and is smallest when orbital $i$ is near integer occupation. As $\lambda_i^{I\sigma} \to \tfrac{1}{2}$, its on-site $+U$ potential $\tfrac{U^I}{2}(1-2\lambda_i^{I\sigma})$ vanishes, so a finite inter-site term is no longer small by comparison and the expansion fails even for $V \ll U$.
\linelabel{degenerate_pointer:start}(The more general form, \cref{eq:explicit_s}, has a vanishing denominator if $V$ couples degenerate levels. The $\lambda_i^{I\sigma} \to \tfrac{1}{2}$ divergence discussed above falls into this category; so too would \emph{e.g.} NiO with an unconventional $V$ applied between like-spin nickel sites, in which case the two sites mix evenly rather than weakly and the hybridized-projector picture breaks down --- see \cref{sec:degenerate}.)\linelabel{degenerate_pointer:end}

Example hybridized orbitals for bulk NiO are shown in Figs.~\ref{fig:all_orbitals}g--\ref{fig:all_orbitals}i. In particular, the spin-minority $e_g$ hybridized projectors gain antibonding character (Figs.~\ref{fig:all_orbitals}j and \ref{fig:all_orbitals}k). Because they are less than half-filled, the $+U$ penalty drives their occupation even further down, depleting antibonding density and strengthening the Ni--O bond --- so applying $+U$ to the hybridized projectors enhances covalency, the opposite of the over-localization caused by $+U$ on the bare projectors. This makes precise the common intuition that $+V$ ``restores covalency'': to first order, $+V$ adds no physics beyond the occupation-curvature penalty of $+U$. What changes is where that penalty is measured: now, it is applied to hybridized projectors that already contain the bond's covalent character, rather than to bare projectors, where the same correction would disrupt that covalency. \gls{dftuv} thus improves on \gls{dftu} wherever the bare projectors misrepresent the covalent character of the subspace --- though whether a deeper principle selects these hybridized projectors is a question to which we will return.

\begin{figure*}[t]
\includegraphics[width=\textwidth]{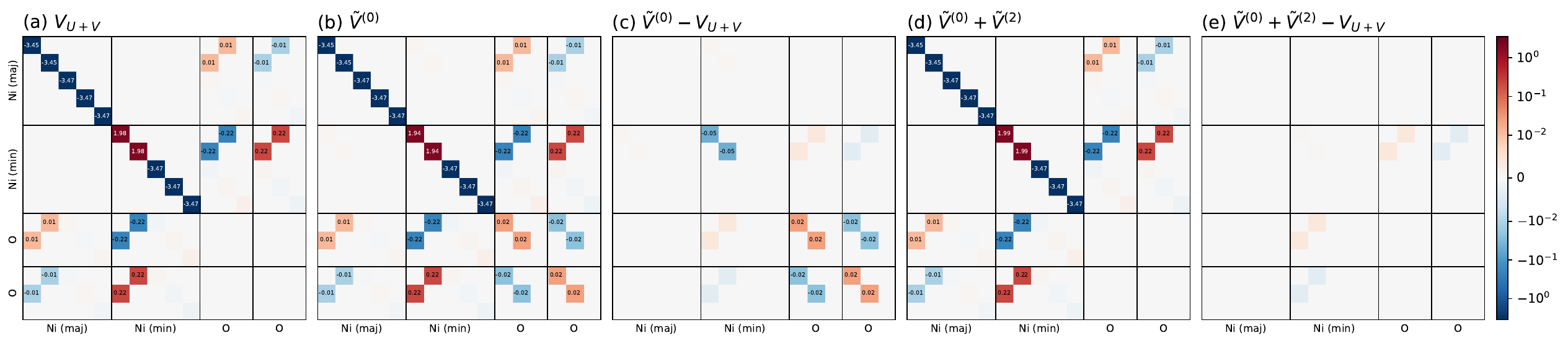}
\caption{The frozen-projector approximation reproduces the exact \gls{dftuv} potential in bulk antiferromagnetic nickel oxide to within a small residual (panels a--c), which can be almost entirely accounted for by a single analytic second-order term (panels d, e). Each panel shows the matrix elements of various potentials in the basis of block-diagonalizing orbitals $\{\varphi_i^{I\sigma}\}$ evaluated at the $\Gamma$-point. The matrices are subdivided by Hubbard site; one can clearly see the five $d$ orbitals of each Ni and the three $p$ orbitals of each O. (The unit cell contains four atoms: two Ni and two O.) Panel (a) shows the matrix elements of the full (\emph{i.e.}\ exact) \gls{dftuv} potential $\braopket{\varphi_i^{I\sigma}}{\hat{V}_U + \hat{V}_V}{\varphi_j^{J\sigma}}$. The on-site blocks of the potential are clearly diagonal in this basis (by design). Panel (b) shows the matrix elements of the approximate potential $\braopket{\varphi_i^{I\sigma}}{\hat{\tilde{V}}^{(0)\sigma}}{\varphi_j^{J\sigma}}$ (\cref{eq:vuv_rewritten}), which, as desired, closely replicate the full \gls{dftuv} potential: the differences between the two are shown in panel (c). Panel (d) shows the matrix elements of the potential $\braopket{\varphi_i^{I\sigma}}{\hat{\tilde{V}}^{(0)\sigma} + \hat{\tilde{V}}^{(2)\sigma}}{\varphi_j^{J\sigma}}$, where $\hat{\tilde{V}}^{(2)\sigma}$ is the higher-order correction (\cref{eq:v_u_hybrid_second_order}). This potential even more closely matches the full \gls{dftuv} potential, as shown by their difference in panel (e). Computational details can be found in \cref{sec:comp_details}.}
\label{fig:potential_elements}
\end{figure*}

\arxivtighten%
As noted above, because the generator $\hat S$ is a functional of the density matrix, the hybridized projectors $\{e^{\hat S}\ket{\varphi_i^{I\sigma}}\}$ are density-matrix-dependent. This is a major departure from standard \gls{dftu} practice, in which the Hubbard projectors are externally specified functions, chosen once and held frozen throughout the calculation --- with only rare exceptions, such as the projector-self-consistent scheme of Ref.~\cite{ORegan2010}. Given that using frozen projectors is standard practice, let us briefly consider this regime. Using frozen projectors does not affect the energy equivalence, because the proof in \hyperref[app:proof]{the Appendix} is an identity for any fixed density matrix: at the density matrix for which $\hat{S}$ is constructed, the frozen hybridized projectors still reproduce the \gls{dftuv} energy. However, the same is not true of the potential. The full potential with density-matrix-dependent projectors corresponds to the functional derivative of the energy with respect to the density matrix:
\arxivsqueeze%
\begin{align}
  \frac{\delta E_U[\hat{S}[\hat\rho]]}{\delta\hat\rho^\sigma} =& \frac{\delta E_U}{\delta\hat\rho^\sigma}\bigg|_{\hat{S}} + \Big\langle\frac{\partial E_U}{\partial\hat{S}},\ \frac{\partial\hat{S}}{\partial\hat\rho^\sigma}\Big\rangle.
  \label{eq:chain_rule}
\end{align}
\arxivsqueeze%
\arxivsqueeze%
Evaluating this derivative (see \cref{sec:potential_breakdown}) gives
\begin{align}
\frac{\delta E_U[\hat{S}[\hat\rho]]}{\delta\hat\rho^\sigma}=& \hat{V}_U^\sigma + \,\hat{V}_V^\sigma + \mathcal{O}(V^2/U),
\end{align}
--- that is, the \gls{dftu} potential applied to hybridized projectors reproduces the \gls{dftuv} potential to first order, as expected. Freezing the projectors amounts to dropping the second term in \cref{eq:chain_rule}. Doing so, one finds that 
if $\hat{T} = 2\hat{S}$ (\emph{i.e.}\ twice the rotation that reproduces the energy), then
\begin{align}
  \hat V^\sigma_{U} + \hat V^\sigma_V
  =& \underbrace{\sum_{Ii} \frac{U^I}{2} \left(1 - 2 \lambda_i^{I\sigma} \right) \ket{\tilde \varphi_i^{I\sigma}}\bra{\tilde\varphi_i^{I\sigma}}}_{\eqqcolon \hat{\tilde{V}}^{(0)\sigma}} + \mathcal{O}(V^2/U),
  \raisetag{1.5\baselineskip}
  \label{eq:vuv_rewritten}
\end{align}
which defines the frozen-projector potential $\hat{\tilde{V}}^{(0)\sigma}$: an on-site Hubbard correction, with no inter-site term, evaluated on the doubly hybridized projectors $\ket{\tilde\varphi_i^{I\sigma}} = e^{2\hat S}\ket{\varphi_i^{I\sigma}}$ (see \cref{sec:potential_breakdown,sec:proof} for the full proof). In other words, we can map \gls{dftuv} to \gls{dftu} with frozen hybridized projectors, but no single choice reproduces both the spectrum (which calls for $e^{2\hat S}$) and the energy ($e^{\hat S}$). This mismatch also manifests itself via the prefactor in \cref{eq:vuv_rewritten}, which is inconsistently built from the bare occupations $\lambda_i^{I\sigma}$ of the original projectors, not from the occupation matrix $n_{ij}^{II\sigma}[2\hat{S}] = \braopket{\tilde\varphi_i^{I\sigma}}{\hat\rho^\sigma}{\tilde\varphi_j^{I\sigma}}$ of the hybridized projectors on which it acts (see \cref{sec:obstruction}).

With or without freezing the projectors, it is important to remember that the proof applies in the perturbative limit $V \ll U$. In practice this limit is only approximately satisfied --- in bulk NiO, typical values are $U \approx 7$~eV and $V \approx 1$~eV~\cite{Campo2010a} --- but the expansion remains well under control. \Cref{fig:potential_elements} compares the exact \gls{dftuv} potential (panel a) with the first-order approximation $\hat{\tilde V}^{(0)\sigma}$ (panel b): the two agree to within a small residual (panel c), and even this residual is predicted analytically by the higher-order term $\hat{\tilde V}^{(2)\sigma}$ (panels d, e) --- albeit at a cost: $\hat{\tilde V}^{(2)\sigma}$ no longer has the on-site $+U$ form (see \cref{sec:higher_order_terms}. Note that the numerical superscript in $\hat{\tilde{V}}^{(n)\sigma}$ counts powers of $\hat{S}$, and the first-order correction $\hat{\tilde{V}}^{(1)\sigma}$ is precisely the term that the defining relation of $\hat{S}$ eliminates. This is why the leading correction to $\hat{\tilde{V}}^{(0)\sigma}$ is the second-order term $\hat{\tilde{V}}^{(2)\sigma}$.)

Our reinterpretation of \gls{dftuv} as \gls{dftu} applied to hybridized density-matrix-dependent projectors has consequences for how $V$ ought to be computed. If $+V$ is really $+U$ on the hybridized projectors, the self-consistent way to parametrize it would be to enforce linearity of the total energy in the occupations of these hybridized projectors, $n^{II}[\hat{S}]$. But that occupation responds through inter-site channels --- both the response \emph{of} the inter-site blocks $\mathbf{n}^{IJ\sigma}$ and the response \emph{to} inter-site perturbations --- that the conventional response matrices do not contain (see \cref{sec:computing_v} for details).
\linelabel{mismatch_compounds:start}In other words, the equivalence we have uncovered does not retroactively justify computing $V$ as the off-diagonal elements of the response matrices used to compute $U$ as is common practice \cite{Campo2010a}; instead, it compounds the pre-existing discrepancy between the site-diagonal occupations the response matrices measure and the inter-site blocks that $E_V$ penalizes.\linelabel{mismatch_compounds:end}

\arxivtighten%
This line of thought raises deeper questions: why might we want to linearize the total energy with respect to the occupation of the hybridized projectors, rather than the original projectors --- or any other choice? Is there an ``optimal'' set of projectors for \gls{dftu}? A broad family of \gls{dftu}-inspired functionals --- \emph{e.g.}\ Koopmans functionals~\cite{Dabo2010,Nguyen2018,Linscott2023}, the localized orbital scaling correction~\cite{Li2018,Su2020}, Wannier-localized optimally-tuned range-separated hybrid functionals~\cite{Wing2021}, and Koopmans-Wannier functionals~\cite{Ma2016} --- corrects quasiparticle energies by imposing piecewise linearity of the total energy with respect to localized orbitals (typically Wannier functions or variants thereof~\cite{Marzari2012,Mahler2025}), and the localized nature of these orbitals is crucial to their success~\cite{Nguyen2018}. 
Why correcting the curvature with respect to localized orbitals should so effectively correct the energies of delocalized quasiparticles deserves further investigation (see, \emph{e.g.}, Ref.~\cite{Ohad2026}).

\linelabel{conclusion:start}To conclude, the inter-site $+V$ corrections of \gls{dftuv} are equivalent --- to first order in $V/U$ --- to an on-site \gls{dftu} correction evaluated on a density-matrix-dependent set of Hubbard projectors hybridized with those of neighboring sites. The role of a $+V$ correction is thus to make the on-site correction aware of covalency. This equivalence is rigorous: when the projectors are allowed to adapt to the density matrix, the single choice of rotation $e^{\hat{S}}$ reproduces both the \gls{dftuv} energy and, through the projectors' own dependence on the density matrix, its potential (\cref{eq:potential_from_energy}).

In contrast, if this redefinition is attempted with \emph{frozen} projectors, as in most current implementations, the equivalence becomes partial: the total energy and the spectrum then call for differently hybridized projectors ($e^{\hat{S}}$ and $e^{2\hat{S}}$, respectively), and no frozen-projector functional reproduces the \gls{dftuv} potential exactly (\cref{sec:obstruction}). Even so, an ordinary \gls{dftu} calculation with hybridized frozen projectors may stand in --- approximately --- for \gls{dftuv}, bringing the far broader ecosystem of \gls{dftu} implementations and property extensions to bear.\linelabel{conclusion:end}

The message to practitioners of \gls{dftu} and \gls{dftuv} is that the choice of projectors, the choice of Hubbard parameters, and the choice of whether to apply inter-site corrections are deeply intertwined and should never be treated as independent degrees of freedom. More deeply --- with \gls{dftuv} now embedded in community codes and automated high-throughput workflows~\cite{Timrov2022,Bastonero2025,Uhrin2025,Beida2026a,Zhang2026}, increasingly the default extension of \gls{dftu} --- this work sharpens the question of what the Hubbard projectors fundamentally are, by revealing that even the established inter-site correction is, at heart, a choice of projectors. Determining how projectors might be chosen optimally is now all the more pressing.

\vspace{2ex}

\begin{acknowledgments}
\noindent The authors thank Maria Andolfatto, Ermanno Botti, Matteo Cococcioni, Nicola Colonna, David O'Regan, and Iurii Timrov for helpful discussions. EL gratefully acknowledges financial support from the Swiss National Science Foundation (grant number 213082). This research was also supported by the NCCR MARVEL, a National Centre of Competence in Research, funded by the Swiss National Science Foundation (grant number 205602).
\end{acknowledgments}

\section*{Author contributions}
\noindent E.L. led the conceptualization, methodology, investigation, and formal analysis, developed the associated software, curated the data and visualizations, and wrote the original draft. A.C. contributed to the investigation, validation, and review of the manuscript. N.M. contributed to the conceptualization of the work and provided funding, and manuscript review and editing.

\section*{Data availability}
\noindent The data supporting the findings of this article are openly available on the Materials Cloud Archive~\cite{LinscottMCA2026}.


\appendix
\phantomsection\label{app:proof}
\section*{Appendix}

\noindent Here we establish the central result of the main text --- that, to first order in $V/U$, \gls{dftuv} is the \gls{dftu} energy functional evaluated on a density-matrix-dependent redefinition of the Hubbard projectors.

The \gls{dftu} and \gls{dftuv} corrections are most transparent in the basis $\{\ket{\varphi_i^{I\sigma}}\}$ that diagonalizes the on-site occupation matrices $\mathbf{n}^{II\sigma}$, with eigenvalues $\lambda_i^{I\sigma}$. The Hubbard projectors are taken orthonormal across sites as well as within them, $\braket{\phi^{I\sigma}_m}{\phi^{J\sigma}_{m'}} = \delta^{IJ}\delta_{mm'}$ (the ``ortho-atomic'' scheme), so that $\{\ket{\varphi_i^{I\sigma}}\}$ is orthonormal, the site projectors below are orthogonal, and the rotations $e^{\hat{T}}$ preserve this structure; non-orthogonal projectors would thread inter-site overlap matrices through the operator algebra that follows. Writing $\Lambda^{IJ\sigma}_{ij} = \braopket{\varphi_i^{I\sigma}}{\hat\rho^\sigma}{\varphi_j^{J\sigma}}$ for the density matrix in this basis (so that $\Lambda^{II\sigma}_{ij} = \lambda_i^{I\sigma}\delta_{ij}$), the on-site potential is diagonal
\begin{equation}
  \hat{V}^\sigma_U = \sum_{Ii} \frac{U^I}{2}\left(1 - 2\lambda_i^{I\sigma}\right)\ket{\varphi^{I\sigma}_i}\bra{\varphi^{I\sigma}_i},
  \label{eq:v_u_diagonalizing}
\end{equation}
while the inter-site potential is purely off-site (recall $V^{II} = 0$),
\begin{equation}
  \hat V^\sigma_V = -\sum_{IJij} V^{IJ}\Lambda^{IJ\sigma}_{ij}\ket{\varphi_i^{I\sigma}}\bra{\varphi_{j}^{J\sigma}}.
  \label{eq:v_v_diagonalizing}
\end{equation}
We start from the \gls{dftu} energy evaluated on an arbitrary unitarily rotated set of projectors $e^{\hat{T}}\ket{\varphi_i^{I\sigma}}$, with $\hat{T}$ any anti-Hermitian generator:
\begin{equation}
  E_U[\hat{T}] = \sum_{I\sigma} \frac{U^I}{2} \trace{\mathbf{n}^{II\sigma}[\hat{T}](1 - \mathbf{n}^{II\sigma}[\hat{T}])},
  \label{eq:e_u_hat}
\end{equation}
where
\begin{align}
n^{II\sigma}_{ij}[\hat{T}]
=& \braopket{\varphi_i^{I\sigma}}{e^{-\hat{T}}\hat\rho^\sigma e^{\hat{T}}}{\varphi_j^{I\sigma}} \nonumber \\
=& \lambda_i^{I\sigma}\delta_{ij} + \delta n^{II\sigma}_{ij}[\hat{T}] + \mathcal{O}(\hat{T}^2),
\label{eq:n_ii_T}
\end{align}
and $\delta n^{II\sigma}_{ij}[\hat{T}] \equiv \braopket{\varphi_i^{I\sigma}}{[\hat\rho^\sigma, \hat{T}]}{\varphi_j^{I\sigma}}$. It follows that
\begin{align}
  & \trace{\mathbf{n}^{II\sigma}[\hat{T}](1 - \mathbf{n}^{II\sigma}[\hat{T}])} \nonumber \\
  =& \sum_i \lambda_i^{I\sigma}(1 - \lambda_i^{I\sigma}) + \sum_i (1 - 2\lambda_i^{I\sigma})\,\delta n^{II\sigma}_{ii}[\hat{T}] + \mathcal{O}(\hat{T}^2).
\end{align}
Inserting this into \cref{eq:e_u_hat} and using \cref{eq:v_u_diagonalizing} gives the first-order energy of \emph{any} such rotation,
\begin{align}
  E_U[\hat{T}]
  &= E_U + \sum_{Ii\sigma} \frac{U^I}{2}(1 - 2\lambda_i^{I\sigma}) \braopket{\varphi_i^{I\sigma}}{[\hat\rho^\sigma,\hat{T}]}{\varphi_i^{I\sigma}} + \mathcal{O}(\hat{T}^2) \nonumber \\
  &= E_U + \sum_\sigma \trace{\hat{V}_U^\sigma\,[\hat\rho^\sigma, \hat{T}]} + \mathcal{O}(\hat{T}^2) 
  \label{eq:e_u_master}
\end{align}
\arxivtighten%
where $E_U$ is the \gls{dftu} energy of the original projectors.

We now ask which specific rotation $\hat{S}$ makes the \gls{dftu} energy of the rotated projectors equal the \gls{dftuv} energy $E_U + E_V$. By \cref{eq:e_u_master} this means that
\begin{equation}
  \sum_\sigma \trace{\hat\rho^\sigma\,[\hat{S},\hat{V}_U^\sigma]} = E_V .
  \label{eq:energy_demand}
\end{equation}
To simplify this condition, we can rewrite the inter-site energy and its potential compactly using the site projector $\hat{P}^{I\sigma} = \sum_m \ket{\phi_m^{I\sigma}}\bra{\phi_m^{I\sigma}}= \sum_i \ket{\varphi_i^{I\sigma}}\bra{\varphi_i^{I\sigma}}$,
\begin{align}
  E_V &= -\sum_{IJ\sigma}\frac{V^{IJ}}{2}\trace{\hat{P}^{I\sigma}\hat\rho^\sigma\hat{P}^{J\sigma}\hat\rho^\sigma}, \label{eq:e_v_projector} \\
  \hat{V}_V^\sigma &= \frac{\delta E_V}{\delta\hat\rho^\sigma} = -\sum_{IJ}V^{IJ}\hat{P}^{I\sigma}\hat\rho^\sigma\hat{P}^{J\sigma}.
  \label{eq:v_v_projector}
\end{align}
It follows that these two quantities can be related via the identity
\begin{equation}
  \sum_\sigma\trace{\hat\rho^\sigma\hat{V}_V^\sigma} = -\sum_{IJ\sigma}V^{IJ}\trace{\hat\rho^\sigma\hat{P}^{I\sigma}\hat\rho^\sigma\hat{P}^{J\sigma}} = 2E_V
  \label{eq:rho_vv}
\end{equation}
(This identity is analogous to the same result for the Hartree energy $2 E_\mathrm{H} = \int \rho\, v_\mathrm{H}\, \mathrm{d}\mathbf{r}$. We obtain the same relation because both energies are quadratic in $\rho$).

Given this identity, the condition of \cref{eq:energy_demand} becomes $\sum_\sigma\trace{\hat\rho^\sigma\big([\hat{S},\hat{V}_U^\sigma] - \tfrac{1}{2}\hat{V}_V^\sigma\big)} = 0$, which is satisfied at every density matrix if
\begin{equation}
  [\hat{S}, \hat{V}_U^\sigma] = \tfrac{1}{2}\hat{V}_V^\sigma .
  \label{eq:s_defining}
\end{equation}

\noindent From this, we can obtain the explicit form
\begin{equation}
  S^{IJ\sigma}_{ij} = \braopket{\varphi_i^{I\sigma}}{\hat{S}}{\varphi_j^{J\sigma}} = \frac{V^{IJ}\Lambda^{IJ\sigma}_{ij}}{U^I(1 - 2\lambda_i^{I\sigma}) - U^J(1 - 2\lambda_j^{J\sigma})},
\end{equation}
which is \cref{eq:explicit_s} of the main text. Note that (a) $S^{II\sigma} = 0$ since $V^{II} = 0$, and (b) because $(\Lambda^{IJ\sigma}_{ij})^* = \Lambda^{JI\sigma}_{ji}$, the generator remains anti-Hermitian --- and the rotation unitary --- even for complex occupations.

Given this choice of rotation, it follows that
\begin{equation}
  E_U[\hat{S}] = E_U + E_V + \mathcal{O}(V^2/U),
  \label{eq:energy_alpha_eq_1}
\end{equation}
precisely as desired. In other words, a \gls{dftu} functional that uses the hybridized projectors $\{e^{\hat{S}}\ket{\varphi^{I\sigma}_i}\}$ yields the same energy as the original \gls{dftuv} functional. Furthermore, because \cref{eq:energy_alpha_eq_1} holds for every $\hat\rho^\sigma$, it may be differentiated:
\begin{equation}
  \frac{\delta E_U[\hat{S}]}{\delta\hat\rho^\sigma} = \hat{V}_U^\sigma + \hat{V}_V^\sigma + \mathcal{O}(V^2/U),
  \label{eq:potential_from_energy}
\end{equation}
which means the potentials are also equivalent.

\nocite{Wang2022b,Giannozzi2009,Giannozzi2017,Carta2025,Perdew2008,vanSetten2018,Hamann2013,Marzari1997,Timrov2018,Timrov2021}

\bibliography{references,footnotes}

\end{document}


\ifrefereelines\linenumbers\fi

\title{Supplemental Material for ``\maintitle''}

\author{Edward Linscott\orcidlink{0000-0002-4967-9873}}
 \email{edward.linscott@psi.ch}
\affiliation{%
 PSI Center for Scientific Computing, Theory and Data, Paul Scherrer Institute, 5232 Villigen PSI, Switzerland}%
\affiliation{%
 National Centre for Computational Design and Discovery of Novel Materials (MARVEL), Paul Scherrer Institute, 5232 Villigen PSI, Switzerland}%
\author{Alberto Carta\orcidlink{0000-0003-0705-0281}}
\affiliation{%
 PSI Center for Scientific Computing, Theory and Data, Paul Scherrer Institute, 5232 Villigen PSI, Switzerland}%
\affiliation{%
 National Centre for Computational Design and Discovery of Novel Materials (MARVEL), Paul Scherrer Institute, 5232 Villigen PSI, Switzerland}%
\author{Nicola Marzari\orcidlink{0000-0002-9764-0199}}%
\affiliation{%
 PSI Center for Scientific Computing, Theory and Data, Paul Scherrer Institute, 5232 Villigen PSI, Switzerland}%
\affiliation{%
 National Centre for Computational Design and Discovery of Novel Materials (MARVEL), Paul Scherrer Institute, 5232 Villigen PSI, Switzerland}%
\affiliation{%
Theory and Simulation of Materials (THEOS), École Polytechnique Fédérale de Lausanne, 1015 Lausanne, Switzerland}%
\affiliation{%
Theory of Condensed Matter, Cavendish Laboratory, University of Cambridge, Cambridge CB3 0US, United Kingdom
\vspace{2ex}}

\date{\today}
\maketitle
\glsunsetall

\section{Hubbard parameters via linear response}
\label{sec:linear_response_details}
\noindent The Hubbard parameters can be computed from first principles using linear-response theory.
\linelabel{spurious_sm:start}The on-site $U^I$ quantifies the local curvature of the total energy with respect to the occupation of Hubbard site $I$ --- the curvature that the $+U$ framework treats as spurious --- and is given by\linelabel{spurious_sm:end}
%
\begin{equation}
U^I = \left(\chi_0^{-1}\right)_{II} - \left(\chi^{-1}\right)_{II}
\label{eq:u_lr}
\end{equation}
where $\chi_{IJ} = \frac{dn^{II}}{dv_\mathrm{ext}^J}$ and $\chi_0 = \frac{dn^{II}}{dv_\mathrm{KS}^J}$ are the interacting and non-interacting responses of the occupation of Hubbard site $I$, $n^{II} = \sum_{m\sigma} n^{II\sigma}_{mm}$, to the infinitesimal external perturbation
\begin{equation}
d\hat{v}^J = dv_\mathrm{ext}^J \sum_{\sigma m} \ket{\phi^{J\sigma}_{m}} \bra{\phi^{J\sigma}_{m}}
\label{eq:dv}
\end{equation}
that shifts the potential of Hubbard site $J$ by the scalar $dv^J_\mathrm{ext}$~\cite{Cococcioni2005a,Timrov2018,footnote1}.

The inter-site Hubbard parameters $V^{IJ}$ are typically computed from the off-diagonal elements of the same response matrices:
\begin{equation}
V^{IJ} = \left(\chi_0^{-1}\right)_{IJ} - \left(\chi^{-1}\right)_{IJ} \qquad \text{for } I \neq J.
\label{eq:v_lr}
\end{equation}

\section{The rotation generator in periodic systems}
\label{sec:periodic}
\noindent For bulk systems, the Hubbard sites exist on a periodic lattice represented with Born--von Kármán periodic boundary conditions. In this framework, the Hubbard overlap matrices and the inter-site coupling terms acquire a dependence on relative lattice-vector displacements (for when sites $I$ and $J$ do not belong to the same cell). This means that the generator (\cref{eq:explicit_s}) becomes
\begin{align}
  S_{ij}^{IJ\sigma}(\mathbf{R} - \mathbf{R}')
  & = \bra{\varphi_i^{\mathbf{R}I\sigma}}\hat{S}\ket{\varphi_j^{\mathbf{R}'J\sigma}} \nonumber \\
  & = \braopket{\varphi_i^{(\mathbf{R} - \mathbf{R}')I\sigma}}{\hat{S}}{\varphi_j^{\mathbf{0}J\sigma}} \nonumber \\
  & = \frac{V^{IJ}(\mathbf{R}-\mathbf{R}') \Lambda^{IJ\sigma}_{ij}(\mathbf{R}-\mathbf{R}')}{U^I(1 - 2\lambda_i^{I\sigma}) - U^J(1 - 2 \lambda_j^{J\sigma})}
\end{align}
(where we have used the fact that $U$ and $\lambda$ are $\mathbf{R}$-independent). It follows that the $k$-space rotation can be written as
\begin{align}
  \mathcal{U}^{IJ\sigma}_{ij}(\mathbf{k}) =& \left[e^{S(\mathbf{k})}\right]_{ij}^{IJ\sigma}
\end{align}
where
\begin{equation}
  S_{ij}^{IJ\sigma}(\mathbf{k}) = \sum_{\mathbf{R}} e^{i\mathbf{k} \cdot \mathbf{R}} S_{ij}^{IJ\sigma}(\mathbf{R})
\end{equation}
(here and throughout, Bloch states are $\ket{\varphi^{\mathbf{k}I\sigma}_i} = N_\mathbf{R}^{-1/2}\sum_\mathbf{R} e^{i\mathbf{k}\cdot\mathbf{R}}\ket{\varphi^{\mathbf{R}I\sigma}_i}$, so that no prefactor appears in the sum over $\mathbf{R}$).
%

\section{Where the perturbative rotation breaks down}
\label{sec:degenerate}

\noindent The explicit form of the generator (\cref{eq:explicit_s}) assumes that the denominator is nonvanishing: it fails whenever $V$ couples two levels for which $U^I(1 - 2\lambda_i^{I\sigma}) = U^J(1 - 2\lambda_j^{J\sigma})$. As stated in the main text, this can arise where $U^J = 0$ and $\lambda_i = \frac{1}{2}$, but also whenever $V$ couples two Hubbard sites whose Hubbard parameters and occupation eigenvalues coincide. Degenerate pairs of this kind are commonplace: the two sites of a spin-symmetric homonuclear dimer, any pair of equivalent sites in a ferromagnet, or --- within a given spin channel --- any pair of equivalent sites on the same spin sublattice of an antiferromagnet. Whether one chooses to apply a $V$ between such a pair is a separate question (to which we will return); here we assume such an inter-site coupling is in place, explain the physics behind the resulting divergence, and sketch how this degenerate case would have to be treated. (We suppress the spin index $\sigma$ throughout.)

Consider first an extreme limit: two equivalent Hubbard sites that share no inter-site occupations (\emph{i.e.} $\Lambda^{IJ} = 0$). Each pair of equivalent block-diagonalizing orbitals $\ket{\varphi_k^{I}}$ and $\ket{\varphi_k^{J}}$ has the same occupation $\lambda_k$, so the two span a degenerate eigenspace: any mixture $\cos\theta \ket{\varphi_k^{I}} + e^{i\phi}\sin\theta \ket{\varphi_k^{J}}$ is again an eigenvector with the same eigenvalue $\lambda_k$. Therefore, mixing these two projectors by \emph{any} amount leaves every on-site occupation matrix, and hence the \gls{dftu} energy, exactly unchanged: \gls{dftu} is completely indifferent to this inter-site projector mixing. (Practically, this indifference to non-locality is completely undesirable, as it departs from our goal of applying linearizing corrections to localized subspaces.)

In the less extreme limit where $\Lambda^{IJ} \neq 0$, this indifference persists but only to first order. The change in the \gls{dftu} energy under a projector rotation with generator $\hat{T}$ is
\begin{align}
  E_U[\hat{T}] - E_U
  &= \trace{\hat{V}_U [\hat\rho, \hat{T}]} + \mathcal{O}(\hat{T}^2) \nonumber \\
  &= \trace{\hat\rho\,[\hat{T},\hat{V}_U]} + \mathcal{O}(\hat{T}^2) \nonumber \\
  &= \sum_{IJij} \left(\varepsilon_j^{J} - \varepsilon_i^{I}\right) \Lambda^{JI}_{ji}\, T^{IJ}_{ij} + \mathcal{O}(\hat{T}^2),
  \label{eq:first_order_rotation}
\end{align}
where $\varepsilon_i^{I} = \frac{U^I}{2}(1 - 2\lambda_i^{I})$ are the eigenvalues of $\hat{V}_U$. The factor $\varepsilon_j^{J} - \varepsilon_i^{I}$ vanishes for degenerate levels, so \gls{dftu} remains blind at first order to any mixing of degenerate projectors. A $+V$ correction between such levels, by contrast, contributes at first order whenever $\Lambda^{IJ}_{ij} \neq 0$. This order-mismatch means that a rotated \gls{dftu} functional cannot possibly absorb the $+V$ energy contribution --- which is why we see the vanishing denominator breakdown of \cref{eq:explicit_s}.

The same factor governs the equation that defines $\hat{S}$. Starting from the defining relation $[\hat{S}, \hat{V}_U] = \frac{1}{2}\hat{V}_V$ (\cref{eq:s_defining}) and taking matrix elements between eigenvectors of $\hat{V}_U$ gives
\begin{equation}
  \left(\varepsilon_j^{J} - \varepsilon_i^{I}\right) S^{IJ}_{ij} = \tfrac{1}{2}\left(V_V\right)^{IJ}_{ij} = -\tfrac{1}{2}V^{IJ}\Lambda^{IJ}_{ij},
  \label{eq:sylvester}
\end{equation}
which determines $S^{IJ}_{ij}$ --- recovering \cref{eq:explicit_s} --- whenever $\varepsilon_i^{I} \neq \varepsilon_j^{J}$, and has no solution at all whenever $\varepsilon_i^{I} = \varepsilon_j^{J}$ while $V^{IJ}\Lambda^{IJ}_{ij} \neq 0$. In this degenerate case, the remedy is to apply degenerate perturbation theory: the zeroth-order projectors must diagonalize $\hat{V}_V$ within each degenerate eigenspace of $\hat{V}_U$. Because \gls{dftu} expresses no preference within that eigenspace, the inter-site coupling dictates the projectors outright --- and however small $V$ may be, the resulting projectors mix into bonding and antibonding combinations carrying equal weight on both sites. Such projectors are delocalized over the site pair --- nothing like site-localized projectors weakly hybridized with their neighbors --- so these channels lie outside the first-order equivalence established in the main text.

In practice, the non-degeneracy condition is comfortably satisfied in many of the settings where $+V$ corrections are typically applied. Transition-metal--ligand couplings link chemically distinct levels, so the denominators in \cref{eq:explicit_s} are of order $U$ so long as the metal levels sit away from half filling; this includes our NiO case study, where $V$ couples only nickel to oxygen and $V^{\text{Ni--Ni}} = 0$ (had it not been, the nickel pairs within each ferromagnetic $(111)$ plane would have been degenerate).

Note also that the divergence as $\lambda_i^{I} \to \tfrac{1}{2}$ encountered in the main text is the same condition in disguise: every level of a ligand site ($U^J = 0$) has $\varepsilon_j^{J} = 0$, and a level of a metal site ($U^I \neq 0$) joins this degenerate set precisely at half filling, meaning that mixing it with the ligands costs no \gls{dftu} energy at first order.

\section{The magnitude of the rotation generator}
\label{sec:s_norm}
\noindent The expansion underpinning the central result of \cref{eq:explicit_s} is controlled by the size of the generator $\hat{S}$ as measured by its spectral norm $\|\hat{S}\|$. Each matrix element of $\hat{S}$ is of order $V\Lambda/U$ for typical magnitudes $V$, $\Lambda$, and $U$ of the inter-site coupling, the inter-site occupations, and the on-site potential differences. However, $\|\hat{S}\|$ is not simply the size of the largest matrix element: a Hubbard orbital couples to all the orbitals of its $V$-coupled neighbors, and these contributions accumulate. One might therefore worry that the sum over neighbors will spoil the smallness of the expansion parameter even when each individual matrix element is small.

This does not happen. To see why, we consider three cases. (Throughout this section we work within a single spin channel, suppressing the spin index $\sigma$.)

\subsection{The single-site and single-orbital limit}
\noindent Consider the simple limit of a single Hubbard site containing a single orbital, coupled to several ligand orbitals that (a) carry no on-site correction of their own and (b) are not coupled to any other Hubbard site. $S^{IJ\sigma}_{ij}$ is non-zero only where $V^{IJ}$ is. Thanks to this structure, the spectral norm of $\hat{S}$ is simply the Euclidean norm of the vector of site--ligand couplings:
\begin{equation}
  \|\hat{S}\| = \frac{V}{U\,|1 - 2\lambda_i|} \sqrt{\sum_{Jj} \big|\Lambda^{JI}_{ji}\big|^2}.
  \label{eq:s_norm_star}
\end{equation}
The neighbor sum thus enters as a root-sum-square of inter-site occupations, which turns out to be bounded. To prove this, insert a complete basis $\{\ket{\chi}\}$ into the expectation value of $\hat{\rho}^2$ in the correlated orbital $\ket{\varphi_i^{I}}$:
\begin{align}
  \braopket{\varphi_i^{I}}{\hat\rho^2}{\varphi_i^{I}}
  &= \sum_{\chi} \big|\braopket{\chi}{\hat\rho}{\varphi_i^{I}}\big|^2 \nonumber \\
  &= \lambda_i^2 + \sum_{Jj} \big|\Lambda^{JI}_{ji}\big|^2
    + \sum_{\chi \notin \{\varphi_i^{I}\} \cup \{\varphi_j^{J} : J, j\}} \big|\braopket{\chi}{\hat\rho}{\varphi_i^{I}}\big|^2 \nonumber \\
  &\geq \lambda_i^2 + \sum_{Jj} \big|\Lambda^{JI}_{ji}\big|^2.
\end{align}
Meanwhile, note that the Kohn-Sham density-matrix operator satisfies $0 \preceq \hat{\rho} \preceq 1$ ($A \preceq B$ denotes the Loewner order for Hermitian operators, meaning that $B - A$ is positive semidefinite; for its eigenvalues this reduces to the familiar inequality $0 \leq f_\nu \leq 1$ for fermions). It follows that $\hat{\rho}^2 \preceq \hat{\rho}$, and so
\begin{equation}
  \braopket{\varphi_i^{I}}{\hat\rho^2}{\varphi_i^{I}} \;\leq\; \braopket{\varphi_i^{I}}{\hat\rho}{\varphi_i^{I}} = \lambda_i.
\end{equation} Therefore
\begin{equation}
  \sum_{Jj} \big|\Lambda^{JI}_{ji}\big|^2 \;\leq\; \lambda_i(1 - \lambda_i) \;\leq\; \tfrac{1}{4}
  \label{eq:covalency_bound}
\end{equation}
--- that is, the total covalency of a correlated orbital is capped by $\lambda_i(1-\lambda_i)$, \emph{independently of the number of ligands}. Adding neighbors redistributes the off-diagonal weight of the density matrix instead of making it larger and larger.

Substituting \cref{eq:covalency_bound} into \cref{eq:s_norm_star} gives the bound
\begin{equation}
  \|\hat{S}\|
  \;\leq\; \frac{V}{U} \frac{\sqrt{\lambda_i(1-\lambda_i)}}{|1 - 2\lambda_i|}
  .
  \label{eq:s_norm_bound}
\end{equation}
Crucially, the expansion parameter is controlled not by a coordination-weighted sum $\sum_J V^{IJ}/U$ but instead simply $V/U$, up to an occupation-dependent factor that is of order unity (unless we are in the region of the known breakdown of the expansion where $\lambda_i \to \tfrac{1}{2}$).

\subsection{The single-site and multi-orbital case}
\noindent The same argument generalizes to the case where site $I$ contains multiple orbitals. Partitioning all the Hubbard orbitals into those that belong to site $I$ and those that do not (denoted $\bar{I}$), let $\rho$ denote the matrix of $\hat{\rho}$ restricted to the incomplete orthonormal basis of Hubbard orbitals, with block form
\begin{equation}
  \rho = \begin{pmatrix} \rho_{II} & \rho_{I\bar{I}} \\ \rho_{\bar{I}I} & \rho_{\bar{I}\bar{I}} \end{pmatrix},
\end{equation}
where $\rho_{II} = \mathrm{diag}(\lambda_i)$ and $(\rho_{I\bar{I}})_{i,Jj} = \Lambda^{IJ\sigma}_{ij}$. Even though the basis is incomplete, the restricted matrix inherits the bounds $0 \preceq \rho \preceq 1$ and thus $\rho^2 \preceq \rho$. Taking the upper-left block of this inequality, we obtain
\begin{align}
  \rho_{I\bar{I}}\,\rho_{\bar{I}I} \;\preceq\; \rho_{II}\,(1 - \rho_{II}),
  \label{eq:covalency_bound_matrix}
\end{align}
which is the matrix generalization of \cref{eq:covalency_bound}. If every correlated orbital is coupled to every ligand orbital with the same $V$ and is subject to the same $U$, the $I\bar{I}$ block of the matrix of $\hat{S}$ is $\tfrac{V}{U} D^{-1} \rho_{I\bar{I}}$ (up to elementwise complex conjugation, which leaves norms unchanged), where $D = \mathrm{diag}(1 - 2\lambda_i)$. Meanwhile, the $II$ and $\bar{I}\bar{I}$ blocks are zero, and thus the matrix norm is given by
\begin{align}
  \|\hat{S}\|^2 
  &= \frac{V^2}{U^2} \big\|D^{-1} \rho_{I\bar{I}}\,\rho_{\bar{I}I}\, D^{-1}\big\| \nonumber \\
  &\leq \frac{V^2}{U^2} \big\|D^{-1} \rho_{II}(1 - \rho_{II})\, D^{-1}\big\| \nonumber \\
  &= \frac{V^2}{U^2}\,\max_i \frac{\lambda_i(1-\lambda_i)}{(1 - 2\lambda_i)^2},
\end{align}
where the inequality applies \cref{eq:covalency_bound_matrix}, and thus
\begin{equation}
  \|\hat{S}\|
  \;\leq\; \frac{V}{U}\,\max_i \frac{\sqrt{\lambda_i(1-\lambda_i)}}{|1 - 2\lambda_i|}
  \label{eq:s_norm_bound_matrix}
\end{equation}
--- \emph{i.e.} the single-orbital bound of \cref{eq:s_norm_bound} evaluated at the worst-case orbital, with no dependence on the number of ligands or correlated orbitals.

\subsection{Extended systems of many sites}
\noindent In extended systems, $\hat{S}$ is block-diagonal in $\mathbf{k}$ (\cref{sec:periodic}) and the expansion parameter is $\max_{\mathbf{k}} \|S(\mathbf{k})\|$. One might hope to apply the bounds derived above at each $\mathbf{k}$-point separately, but this is not possible: $S(\mathbf{k})$ is constructed from only those density-matrix elements that connect $V$-coupled neighbors, and a subset of elements does not itself constitute a density matrix, so the positivity argument that underpins \cref{eq:covalency_bound,eq:covalency_bound_matrix} does not apply. The covalency bound does, however, control the Brillouin-zone \emph{average}: by Parseval's theorem,
\begin{align}
  \frac{1}{N_\mathbf{k}} \sum_\mathbf{k} \sum_{Jj} \big|S^{IJ\sigma}_{ij}(\mathbf{k})\big|^2
  &= \sum_{Jj\mathbf{R}} \big|S^{IJ\sigma}_{ij}(\mathbf{R})\big|^2 \nonumber \\
  &\leq\; \frac{V^2}{U^2} \frac{\lambda_i(1-\lambda_i)}{(1-2\lambda_i)^2},
  \label{eq:s_norm_parseval}
\end{align}
where the sums run over the couplings of a single correlated orbital $i$ to every ligand orbital $(J, j)$, and the final inequality is \cref{eq:covalency_bound} applied to the lattice-summed covalency. The single-orbital bound of \cref{eq:s_norm_bound} therefore controls the average size of the rows of $S(\mathbf{k})$ over the Brillouin zone.

As it turns out, for bulk NiO the norm is also nearly independent of $\mathbf{k}$: $\|S(\mathbf{k})\|$ ranges between $0.0778$ and $0.0782$ across the Brillouin zone, tight against the single-orbital bound (\cref{eq:s_norm_bound}) of 0.080 for the minority-spin $e_g$ orbitals ($V/U = 0.77/6.95 \approx 0.11$; $\lambda_{e_g} = 0.214$). This is also consistent with the first-order residual observed in \cref{fig:potential_elements}, whose relative magnitude of $\sim 1\%$ matches the expected $\mathcal{O}(\|\hat{S}\|^2)$.

\section{Computational details}
\label{sec:comp_details}
\noindent The illustrative Hubbard projectors and potentials of NiO presented in \cref{fig:all_orbitals,fig:potential_elements} were obtained from \gls{dftuv} calculations of antiferromagnetic NiO performed with \texttt{Quantum ESPRESSO} \cite{Giannozzi2009,Giannozzi2017}. NiO was modeled in its rock-salt structure with the experimentally observed type-II antiferromagnetic ordering (alternating ferromagnetic $(111)$ planes of opposite spin), using the corresponding four-atom rhombohedral magnetic cell (two Ni and two O; conventional cubic lattice parameter $a = 4.09$\,\AA). The calculations were spin-polarized and used the PBEsol exchange-correlation functional \cite{Perdew2008}, norm-conserving pseudopotentials from the \texttt{PseudoDojo} library (version 0.4.1; standard accuracy) \cite{vanSetten2018,Hamann2013}, wavefunction and charge-density kinetic-energy cutoffs of $70$ and $280$\,Ry, respectively, an $8\times 8\times 8$ $k$-point grid, and cold smearing \cite{Marzari1997} of width $0.02$\,Ry. The Hubbard projectors were the Ni $3d$ and O $2p$ orbitals, taken from the full set of pseudo-atomic orbitals provided by the pseudopotentials after L\"owdin orthogonalization of that entire set (the ``ortho-atomic'' scheme as implemented in \texttt{Quantum ESPRESSO}).

The Hubbard parameters were computed from first principles using density-functional perturbation theory, as implemented in the \texttt{HP} code of \texttt{Quantum ESPRESSO} \cite{Timrov2018,Timrov2021,Timrov2022}, with a $2\times 2\times 2$ $q$-point grid. This yielded an on-site $U^\text{Ni} = 6.95$\,eV on each Ni $3d$ site and an inter-site $V^\text{Ni-O} = 0.77$\,eV between each Ni $3d$ site and the $2p$ orbitals of its six neighboring oxygens (the oxygen sites themselves carry no on-site $U$).

The block-diagonalizing projectors $\ket{\varphi_i^{I\sigma}}$, the hybridized projectors $\ket{\tilde\varphi_i^{I\sigma}}$, and the potential matrices of \cref{fig:potential_elements,fig:potential_kpoints} were obtained by post-processing the resulting occupation and potential matrices. To evaluate the potential of the hybridized (rotated) projectors, which \texttt{pw.x} cannot construct natively, the rotated orbitals were supplied as custom localized Hubbard projectors through the \texttt{wannier2pw} interface \cite{Carta2025}.

The full input and output files from these calculations, together with the post-processing scripts, can be found on the Materials Cloud Archive \cite{LinscottMCA2026}.

\Cref{fig:all_orbitals} was generated using \texttt{Beautiful Atoms} \cite{Wang2022b}.

\section{Breaking down the contributions to the potential}
\label{sec:potential_breakdown}
\noindent Let us reconsider \cref{eq:potential_from_energy}, separating the explicit and implicit (via $\hat{S}$) density-matrix-dependence of the energy functional. For reasons that will become clear later, we introduce a continuous parameter $\alpha$ that scales the generator $\hat{S}$, so that the energy of the rotated projectors is $E_U[\alpha\hat{S}[\hat\rho]]$. The functional derivative of this energy with respect to the density matrix is then
\begin{equation}
  \frac{\delta E_U[\alpha\hat{S}[\hat\rho]]}{\delta\hat\rho^\sigma} = \underbrace{\frac{\delta E_U}{\delta\hat\rho^\sigma}\bigg|_{\alpha\hat{S}}}_{\text{frozen}} + \underbrace{\Big\langle\frac{\partial E_U}{\partial\hat{S}},\ \frac{\partial\hat{S}}{\partial\hat\rho^\sigma}\Big\rangle}_{\text{projector response}}.
  \label{eq:chain_rule_sm}
\end{equation}

\subsection{The frozen contribution}
\noindent Because \cref{eq:e_u_master} is linear in $\hat{T}$, we have
\begin{equation}
  E_U[\alpha\hat{S}] = E_U + {\alpha}\sum_\sigma\trace{\hat{V}^\sigma_U [\hat\rho^\sigma, \hat{S}]} + \mathcal{O}(\alpha^2 \hat{S}^2).
  \label{eq:energy_alpha}
\end{equation}
and thus the frozen contribution to the potential is
\begin{align}
  \delta_{\hat{\rho}^\sigma} E_U[\alpha\hat{S}] = & \trace{\hat{V}_U^\sigma \delta\hat{\rho}^\sigma} + {\alpha} \trace{\delta \hat{V}^\sigma_U [\hat{\rho}^\sigma, \hat{S}]}
  \nonumber \\ &
  + \frac{\alpha}{2} \trace{\hat{V}^\sigma_V \delta \hat{\rho}^\sigma} + \mathcal{O}(\alpha^2\hat{S}^2)
  \label{eq:delta_energy_alpha_frozen}
\end{align}
where we have taken advantage of the cyclic property of the trace and the definition of $\hat{S}$ to simplify the final term. The second term can be easily evaluated if we write the Hubbard potential in terms of the fixed atomic site projector operators $\hat{P}^{I\sigma}$:
\begin{align}
  \hat{V}_U^\sigma &= \sum_I \frac{U^I}{2}\,\hat{P}^{I\sigma}(1 - 2\hat\rho^\sigma)\hat{P}^{I\sigma}
\end{align}
so that, since $\hat{V}_U^\sigma$ is affine in $\hat\rho^\sigma$, its first variation is simply
\begin{align}
  \delta\hat{V}_U^\sigma = -\sum_K U^K\,\hat{P}^{K\sigma}\,\delta\hat\rho^\sigma\,\hat{P}^{K\sigma};
  \label{eq:dvu}
\end{align}
and, recalling that the change in the occupation matrix induced by a rotation is given by $\delta n^{II\sigma}_{ij}[\hat{T}] = n^{II\sigma}_{ij}[\hat{T}] - n^{II\sigma}_{ij}[\hat{I}]$ = $\braopket{\varphi_i^{I\sigma}}{[\hat\rho^\sigma, \hat{T}]}{\varphi_j^{I\sigma}}$, we obtain
\begin{align}
  \delta \hat{V}^\sigma_U [\hat{\rho}^\sigma, \hat{S}]
  = & - \underbrace{\sum_{Iij} U^I \delta n^{II\sigma}_{ij}[\hat{S}] \ket{\varphi_i^{I\sigma}}\bra{\varphi_j^{I\sigma}}}_{\eqqcolon \tfrac{1}{2}\hat{W}^\sigma}\delta\hat\rho^\sigma
  \label{eq:W_definition}
\end{align}
and thus
\begin{align}
  \frac{\delta E_U}{\delta \hat{\rho}^\sigma}\bigg|_{\alpha\hat{S}}
  =& \hat{V}_U^\sigma + \frac{\alpha}{2}(\hat{V}_V^\sigma - \hat{W}^\sigma) + \mathcal{O}(V^2/U)
  \label{eq:frozen_grad}
\end{align}
This is the entire potential for \gls{dftu} applied to hybridized, \emph{frozen} projectors. Let us consider what choice of $\alpha$ might reproduce the \gls{dftuv} potential. $\hat{V}_U^\sigma$ and $\hat{W}^\sigma$ are block-diagonal operators, while $\hat{V}_V^\sigma$ is purely off-diagonal. This makes it impossible to recover the \gls{dftuv} potential, because matching the off-diagonal blocks requires $\alpha = 2$, in which case
\begin{align}
  \frac{\delta E_U}{\delta \hat{\rho}^\sigma}\bigg|_{\alpha\hat{S}}
  =& \hat{V}_U^\sigma + \hat{V}_V^\sigma - \hat{W}^\sigma \neq \hat{V}_U^\sigma + \hat{V}_V^\sigma .
\end{align}
With \emph{frozen} projectors this is the whole potential, short of the \gls{dftuv} potential by the on-site operator $\hat{W}^\sigma$.

\subsection{The projector response}
\noindent Let us now consider the second term in \cref{eq:chain_rule_sm}, which accounts for the implicit density-matrix-dependence of the energy through the hybridized projectors. To leading order in $V/U$ we obtain (the factor of $\alpha$ arising because the generator enters as $\alpha\hat{S}$)
\begin{equation}
  \delta_{\hat{S}} E_U
  = \alpha\sum_\sigma\trace{\hat\rho^\sigma[\delta\hat{S},\hat{V}_U^\sigma]}.
  \label{eq:force}
\end{equation}
We can avoid evaluating $\delta\hat{S}$ directly by differentiating the defining relation of $\hat{S}$ (\cref{eq:s_defining}) to obtain the identity
\begin{equation}
  [\delta\hat{S},\hat{V}_U^\sigma] = \tfrac{1}{2}\delta\hat{V}_V^\sigma - [\hat{S},\delta\hat{V}_U^\sigma].
  \label{eq:response}
\end{equation}

To resolve the first term we need the off-site potential and its variation with respect to the density matrix, which are given by
\begin{align}
  \hat{V}_V^\sigma &= -\sum_{IJ}V^{IJ}\,\hat{P}^{I\sigma}\hat\rho^\sigma\hat{P}^{J\sigma}
\end{align}
and
\begin{align}
  \delta\hat{V}_V^\sigma &= -\sum_{IJ}V^{IJ}\,\hat{P}^{I\sigma}\,\delta\hat\rho^\sigma\,\hat{P}^{J\sigma}
  \label{eq:dvv}
\end{align}
and thus
\begin{align}
  \trace{\hat{\rho}^\sigma\delta\hat{V}_V^\sigma}
  &= -\sum_{IJ}V^{IJ}\trace{\hat{\rho}^\sigma\hat{P}^{I\sigma}\delta\hat\rho^\sigma\hat{P}^{J\sigma}} \nonumber \\
  &= \trace{\hat{V}_V^\sigma\delta\hat\rho^\sigma}.
  \label{eq:response_1}
\end{align}

Meanwhile, for the second term of \cref{eq:response} we can use \cref{eq:W_definition} to obtain
\begin{align}
  \trace{\hat{\rho}^\sigma[\hat{S},\delta\hat{V}_U^\sigma]}
  &= \trace{\delta\hat{V}_U^\sigma[\hat{\rho}^\sigma,\hat{S}]}
  = -\frac{1}{2}\trace{\hat{W}^\sigma\delta\hat\rho^\sigma}
  \label{eq:response_2}
\end{align}

Combining \cref{eq:force,eq:response,eq:response_1,eq:response_2} gives the projector-response contribution to the potential,
\begin{equation}
  \Big\langle\frac{\partial E_U}{\partial\hat{S}},\ \frac{\partial\hat{S}}{\partial\hat\rho^\sigma}\Big\rangle = \frac{\alpha}{2}\left(\hat{V}_V^\sigma + \hat{W}^\sigma\right) + \mathcal{O}(V^2/U).
  \label{eq:pulay_grad}
\end{equation}
Adding the frozen \cref{eq:frozen_grad} and projector-response \cref{eq:pulay_grad} gradients, the $\hat{W}^\sigma$ term is canceled and the missing half of $\hat{V}_V^\sigma$ is restored,
\begin{equation}
  \frac{\delta E_U[\alpha\hat{S}(\hat\rho)]}{\delta\hat\rho^\sigma} = \hat{V}_U^\sigma + \alpha\,\hat{V}_V^\sigma + \mathcal{O}(V^2/U),
\end{equation}
which matches the \gls{dftuv} potential when $\alpha = 1$, in agreement with \cref{eq:potential_from_energy}.
\vspace{1ex}

\section{A direct derivation of the frozen projectors}
\label{sec:proof}

\noindent If we treat the Hubbard projectors as being frozen, a much more direct derivation of the projectors is possible by directly diagonalizing the corrective potential $\hat{V}_U^\sigma + \hat{V}_V^\sigma$ to first order in $V/U$.

Consider the rotated projectors $\ket{\tilde{\varphi}_i^{I\sigma}} = e^{\hat{T}}\ket{\varphi_i^{I\sigma}}$. These span the Hubbard manifold, so we may expand the corrective potential in this rotated basis:
\begin{equation}
  \hat{V}_U^\sigma + \hat{V}_V^\sigma = \sum_{IJij} \ket{\tilde{\varphi}_i^{I\sigma}}\bra{\tilde{\varphi}_i^{I\sigma}} \hat{V}_U^\sigma + \hat{V}_V^\sigma \ket{\tilde{\varphi}_j^{J\sigma}}\bra{\tilde{\varphi}_j^{J\sigma}}.
\end{equation}
with the individual matrix elements given by
\begin{align}
  \bra{\tilde{\varphi}_i^{I\sigma}} \hat{V}_U^\sigma + \hat{V}_V^\sigma \ket{\tilde{\varphi}_j^{J\sigma}}
  = \bra{{\varphi}_i^{I\sigma}} e^{-\hat{T}}(\hat{V}_U^\sigma + \hat{V}_V^\sigma) e^{\hat{T}} \ket{{\varphi}_j^{J\sigma}}
\end{align}
If we treat the inter-site interactions and the rotation perturbatively, we can expand these matrix elements via
\begin{equation}
  e^{-\hat{T}} \hat V_U^\sigma e^{\hat{T}} + e^{-\hat{T}} \hat V_V^\sigma e^{\hat{T}} \approx \hat{V}_U^\sigma + (\hat{V}_V^\sigma + [\hat{V}_U^\sigma, \hat{T}]) + \mathcal{O}(V^2/U)
\end{equation}
and thus $\hat{V}^\sigma_U + \hat{V}^\sigma_V$ is diagonalized to first order by $\{\tilde{\varphi}_i^{I\sigma}\}$ provided $\hat{V}_V^\sigma + [\hat{V}_U^\sigma, \hat{T}] = 0$ --- which is precisely the defining relation \cref{eq:s_defining} of the generator $\hat{S}$ if $\hat{T} = 2 \hat{S}$, whose explicit form is \cref{eq:explicit_s}. The doubly hybridized projectors $e^{2\hat{S}}\ket{\varphi_i^{I\sigma}}$ therefore block-diagonalize the full corrective potential, as anticipated in the Appendix of the main text.
Using this definition for $\hat{S}$, the rotated orbitals are given by
\begin{widetext}
\begin{equation}
  \ket{\tilde{\varphi}_i^{I\sigma}} = e^{2\hat{S}}\ket{\varphi_i^{I\sigma}} \approx \ket{\varphi_i^{I\sigma}} + \sum_{J} \sum_j \frac{2V^{IJ} \Lambda^{JI\sigma}_{ji}}{U^J(1 - 2\lambda_j^{J\sigma}) - U^I(1 - 2 \lambda_i^{I\sigma})} \ket{\varphi_j^{J\sigma}}.
\end{equation}
Furthermore, given this definition, the total corrective potential can be rewritten as
\begin{align}
  \hat V^\sigma_U + \hat V^\sigma_V
  = &
  \sum_{Ii} \bra{\tilde{\varphi}^{I\sigma}_i} \hat V^\sigma_U + \hat V^\sigma_V \ket{\tilde{\varphi}^{I\sigma}_i} \ket{\tilde{\varphi}^{I\sigma}_i}\bra{\tilde{\varphi}^{I\sigma}_i} 
  = 
  \sum_{Ii} \frac{U^I}{2} (1 - 2 \lambda_i^{I\sigma}) \ket{\tilde{\varphi}^{I\sigma}_i}\bra{\tilde{\varphi}^{I\sigma}_i} + \mathcal{O}(V^2/U)
\end{align}
\vspace{8ex}
\end{widetext}

\noindent In other words, to first order in $V/U$, a combined $+U+V$ correction is equivalent to
\begin{equation}
  \hat{\tilde{V}}^{(0)\sigma} = \sum_{Ii} \frac{U^I}{2} (1 - 2 \lambda_i^{I\sigma}) \ket{\tilde{\varphi}^{I\sigma}_i}\bra{\tilde{\varphi}^{I\sigma}_i}
  \label{eq:v_u_hybrid}
\end{equation}
\emph{i.e.}\ a $+U$-type correction constructed using the rotated Hubbard projectors (albeit using the original orbitals to define $\lambda_i^{I\sigma}$). This is \cref{eq:vuv_rewritten} in the main text.

Note that the prefactor in \cref{eq:vuv_rewritten} uses the original occupations $\lambda_i^{I\sigma}$ rather than those of the hybridized projectors $\tilde{\lambda}_i^{I\sigma} = \braopket{\tilde\varphi_i^{I\sigma}}{\hat\rho^\sigma}{\tilde\varphi_i^{I\sigma}}$. This makes \cref{eq:vuv_rewritten} an operator identity rather than the functional derivative of an on-site Hubbard energy; the mismatch has its origin in the residual $\hat{W}^\sigma$ term of the preceding section, and it is not a counterexample to the proof provided there.

\section{The frozen-projector obstruction}
\label{sec:obstruction}

\noindent Among other things, the previous section showed that for a generator $\hat{T} = \alpha \hat{S}$ it is not possible to reproduce the \gls{dftuv} potential to first order. A natural follow-up is to consider if a more general rotation might succeed. (Throughout this section, we suppress the spin index $\sigma$ for brevity.)

\begin{figure}[t]
\includegraphics[width=\columnwidth]{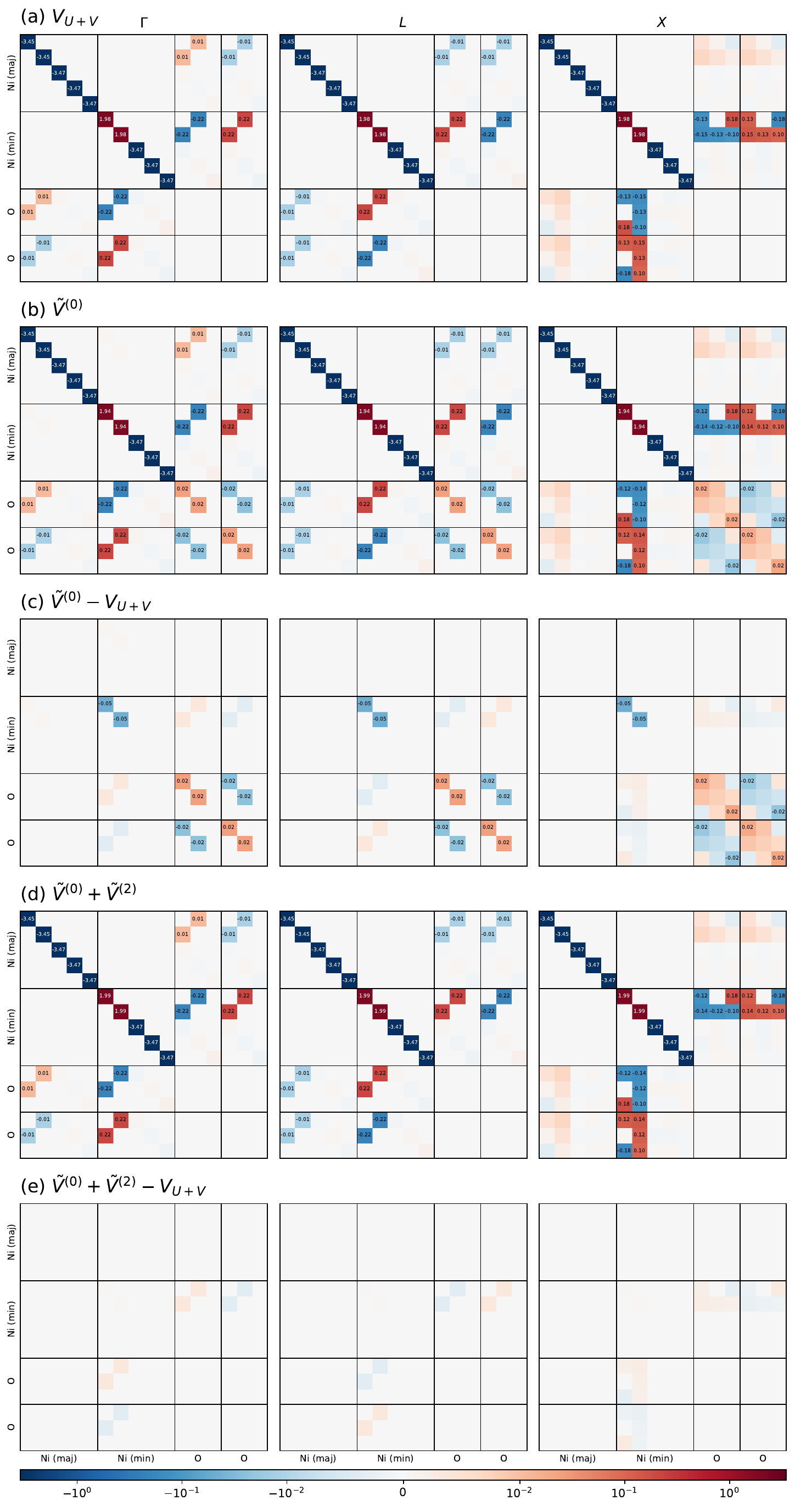}
\caption{An expanded version of \cref{fig:potential_elements} from the main text, evaluated at three high-symmetry $k$-points ($\Gamma$, $L$, and $X$; one per column).}
\label{fig:potential_kpoints}
\end{figure}

Consider the genuine \gls{dftu} functional built on a frozen set of hybridized projectors $e^{\hat{T}}\ket{\varphi_i^{I}}$, for an arbitrary anti-Hermitian generator $\hat{T}$, and evaluated with the \emph{self-consistent} occupations of those projectors --- in contrast to the bare $\lambda_i^I$ carried by the operator $\hat{\tilde{V}}^{(0)}$ (\cref{eq:vuv_rewritten}). Differentiating, its potential is $\hat{V}_U + [\hat{T},\hat{V}_U] - \hat{W}[\hat{T}]$, where $\hat{W}[\hat{T}] = \sum_I U^I \hat{P}^{I}[\hat\rho,\hat{T}]\hat{P}^{I}$. We ask whether any $\hat{T}$ makes this equal the \gls{dftuv} potential, \emph{i.e.}
\begin{equation}
  [\hat{T},\hat{V}_U] - \hat{W}[\hat{T}] = \hat{V}_V .
  \label{eq:obstruction_condition}
\end{equation}
Taking matrix elements, the equation splits by block. For the off-site blocks ($I \neq J$), $\hat{W}$ does not contribute, while
\begin{equation}
  ([\hat{T},\hat{V}_U])^{IJ}_{ij} = \left(\tfrac{U^J}{2}(1 - 2\lambda_j^J) - \tfrac{U^I}{2}(1 - 2\lambda_i^I)\right)T^{IJ}_{ij}.
\end{equation}
Consequently, \cref{eq:obstruction_condition} reduces to
\begin{equation}
  \left(\tfrac{U^J}{2}(1 - 2\lambda_j^J) - \tfrac{U^I}{2}(1 - 2\lambda_i^I)\right)T^{IJ}_{ij} = -V^{IJ}\Lambda^{IJ}_{ij},
\end{equation}
fixing the off-diagonal blocks of $\hat{T}$ to the generator of the hybridizing rotation, $\hat{T} = 2\hat{S}$ (twice \cref{eq:explicit_s}).

For the on-site blocks ($I = J$), $\hat{V}_V$ does not contribute while
\begin{equation}
  ([\hat{T},\hat{V}_U])^{II}_{ij} = U^I \left(\lambda_i^I - \lambda_j^I\right)T^{II}_{ij},
  \label{eq:commutator_ii}
\end{equation}
and
\begin{align}
  \frac{(\hat{W}[T])^{II}_{ij}}{U^I}
  =& \braopket{\varphi_i^I}{[\hat\rho,\hat{T}]}{\varphi_j^I} \nonumber \\*
  =& 
    (\lambda^I_i - \lambda^I_j) T^{II}_{ij}
    + \sum_{K\neq I,k} \left(\Lambda^{IK}_{ik}T^{KI}_{kj} - T^{IK}_{ik}\Lambda^{KI}_{kj}
  \right).
  \label{eq:w_ii}
\end{align}
Inserting \cref{eq:commutator_ii,eq:w_ii} into \cref{eq:obstruction_condition} gives the on-site condition
\begin{equation}
  \sum_{K\neq I,k} \left(\Lambda^{IK}_{ik}T^{KI}_{kj} - T^{IK}_{ik}\Lambda^{KI}_{kj}\right) = 0
\end{equation}
where the on-site terms cancel exactly and we are left with a term that corresponds to the occupation change induced by the (already fixed) inter-site rotation. The unknown $T^{II}_{ij}$ drops out, and the remaining constraint is generically violated whenever the $+V$ correction does anything --- its diagonal ($i = j$) is the on-site occupation change on hybridization, $\tilde\lambda_i^I - \lambda_i^I \neq 0$. \Cref{eq:obstruction_condition} therefore has no solution and the proof is complete: no frozen rotation can reproduce the \gls{dftuv} potential.

The cancellation of $T^{II}$ is not accidental: it is the rotational invariance of the \gls{dftu} energy. On-site rotations cannot change the corresponding potential, while the inter-site rotations that can are entirely consumed in matching $\hat{V}_V$, leaving no additional degree of freedom to satisfy \cref{eq:obstruction_condition}.
\section{Higher-order corrections}
\label{sec:higher_order_terms}
\noindent Note that \cref{eq:v_u_hybrid} is only accurate to first order in $V/U$. We can also expand to higher order:
\begin{widetext}
\begin{align}
  \hat V^\sigma_U + \hat V^\sigma_V
  = &
  \sum_{IJij} \braopket{{\varphi}^{I\sigma}_i}{\hat V^\sigma_U + [\hat{V}_V,\hat{S}]}{{\varphi}^{J\sigma}_j} \ket{\tilde{\varphi}^{I\sigma}_i}\bra{\tilde{\varphi}^{J\sigma}_j}
  + \mathcal{O}(V^3/U^2)
  \nonumber \\
  = & 
  \sum_{Ii} \frac{U^I}{2} (1 - 2 \lambda_i^{I\sigma}) \ket{\tilde{\varphi}^{I\sigma}_i}\bra{\tilde{\varphi}^{I\sigma}_i}
  - \sum_{IJij} \sum_{Kk} \left(V^{IK} \Lambda^{IK\sigma}_{ik} S_{kj}^{KJ\sigma} - S^{IK\sigma}_{ik}V^{KJ}\Lambda^{KJ\sigma}_{kj}\right)\ket{\tilde{\varphi}^{I\sigma}_i} \bra{\tilde{\varphi}^{J\sigma}_j}
  + \mathcal{O}(V^3/U^2)
\end{align}
\emph{i.e.}\ we have the second-order correction
\begin{equation}
  \hat{\tilde{V}}^{(2)\sigma} =
  - \sum_{IJij} \sum_{Kk} \left(V^{IK} \Lambda^{IK\sigma}_{ik} S_{kj}^{KJ\sigma} - S^{IK\sigma}_{ik}V^{KJ}\Lambda^{KJ\sigma}_{kj}\right)\ket{\tilde{\varphi}^{I\sigma}_i} \bra{\tilde{\varphi}^{J\sigma}_j}
\label{eq:v_u_hybrid_second_order}
\end{equation}
\end{widetext}
%
%
In practical calculations employing \gls{dftuv}, we often have a system with two sublattices of Hubbard sites (\emph{e.g.}\ transition metals and oxygens) and $V^{IJ}$ is bipartite (\emph{i.e.}\ only non-zero between a transition metal and an oxygen). In such cases, the second-order correction is block-diagonal in the rotated basis since $V^{IK}S^{KJ}$ is zero unless $I$ and $J$ belong to the same sublattice.

Note that this second-order term is no longer a Dudarev-type $+U$ correction. The first-order term $\hat{\tilde{V}}^{(0)\sigma}$ retained the on-site, projector-diagonal form of $+U$, simply evaluated on the hybridized projectors, whereas $\hat{\tilde{V}}^{(2)\sigma}$ instead couples projectors on distinct Hubbard sites $I \neq J$, with a strength set by $V$ acting through an intermediate site $K$. This coupling could always be removed by a further rotation, $\ket{\tilde{\tilde{\varphi}}_i^{I\sigma}} = e^{\hat{S}'}\ket{\tilde\varphi_i^{I\sigma}}$, so the potential keeps the diagonal operator form to all orders. What it lacks is the structure of a $+U$ \emph{functional}, in which the eigenvalue on each projector is of the form $\tfrac{U^I}{2}(1 - 2\lambda^{I\sigma}_i)$. That the first-order potential nonetheless takes exactly this $+U$ form is not automatic: although any perturbation can be diagonalized in some rotated basis, here both the basis and the resulting potential have an explicit, closed form, and the potential is $+U$-like to lowest order --- which is what gives the $+V$ correction a clear chemical interpretation.

The quality of the frozen-projector approximation is not special to the $\Gamma$-point shown in \cref{fig:potential_elements} of the main text. \Cref{fig:potential_kpoints} repeats that comparison at three different $k$-points, with the same close agreement throughout the Brillouin zone.

\section{Implications for how \texorpdfstring{$V$}{V} is computed}
\label{sec:computing_v}
\noindent The equivalence between \gls{dftuv} and \gls{dftu} with redefined projectors bears on how the inter-site parameters $V^{IJ}$ should be computed. Suppose we are performing a \gls{dftuv} calculation, knowing that the inter-site corrections are equivalent to rotating the Hubbard projectors, $\ket{\varphi_i^{I\sigma}} \to e^{\hat{S}}\ket{\varphi_i^{I\sigma}}$. One might hypothesize that the correct linear-response strategy is to impose that the total energy be linear in the occupation of these \emph{hybridized} projectors, \emph{i.e.}\ in
\begin{align}
  n^{II}[\hat{S}] = & \sum_{i\sigma} \braopket{\varphi_i^{I\sigma}}{e^{-\hat{S}}\hat \rho^\sigma e^{\hat{S}}}{\varphi_i^{I\sigma}} \nonumber \\
  = & n^{II}[\hat{I}] + \sum_{i\sigma}\braopket{\varphi_i^{I\sigma}}{[\hat \rho^\sigma, \hat{S}]}{\varphi_i^{I\sigma}} + \mathcal{O}\left(\tfrac{V^2}{U^2}\right).
  \label{eq:n_tilde}
\end{align}
In this case, one would measure the total-energy curvature via response to perturbations of the form
\begin{align}
  dv^J_\mathrm{ext} \sum_{j\sigma} e^{\hat{S}}\ket{\varphi_j^{J\sigma}} \bra{\varphi_j^{J\sigma}}e^{-\hat{S}},
\end{align}
which would give rise to response matrices $\chi_{IJ}[\hat{S}] = \frac{d n^{II}[\hat{S}]}{d v^J_\mathrm{ext}}$, and one could follow the conventional linear-response strategy to calculate $U$ for the hybridized subspace. However, these new response matrices cannot be rewritten solely in terms of response matrices of the form $\chi_{IJ} = \frac{d n^{II}}{d v^J_\mathrm{ext}}$ (\emph{i.e.}\ those for the original set of projectors). Instead, they also contain the response of the inter-site blocks $\mathbf{n}^{IJ\sigma}$ and the response to inter-site perturbations $\ket{\varphi_i^{I \sigma}}\bra{\varphi_j^{J \sigma}}$ that arise from the commutator in \cref{eq:n_tilde}. Terms of this form are absent from the established linear-response formalism (\cref{eq:u_lr,eq:v_lr}), which features only the response of the site-diagonal and site-summed occupations $n^{II}$ to site-diagonal perturbations. As such, the interpretation of \gls{dftuv} as \gls{dftu} with redefined projectors does not retrospectively justify computing $V$ via \cref{eq:v_lr}: this recipe does not yield a \gls{dftuv} functional equivalent to a \gls{dftu} functional that linearizes the total energy with respect to the occupation of the hybridized orbitals (not to mention that one would also obtain different values for the on-site $U$ parameters).

\linelabel{mismatch_sm:start}It is worth separating this from a discrepancy that is already present in \gls{dftuv} as it is conventionally applied, independently of any reinterpretation. For the on-site parameters, the recipe and the functional address the same derivative: $\partial^2 E_U / \partial (\lambda_i^{I\sigma})^2 = -U^I$ is the curvature that \cref{eq:u_lr} measures (modulo averaging over spins and summing over the orbitals of the site). For the inter-site parameters they do not. \Cref{eq:v_lr} is built from $\chi_{IJ} = \frac{d n^{II}}{d v^J_\mathrm{ext}}$, the response of one site-diagonal occupation to a perturbation of another, and so measures the curvature $\frac{\partial^2 E}{\partial n^{II} \partial n^{JJ}}$. The inter-site energy, meanwhile, is quadratic in the \emph{off-diagonal} blocks, so the curvature it introduces is $\frac{\partial^2 E_V}{\partial n^{IJ\sigma}_{mm'} \partial n^{JI\sigma}_{m'm}} = -V^{IJ}$.
The parameter that is computed and the parameter that appears in the functional are therefore conjugate to different occupations. This has been noted before \cite{Bastonero2025}, where the resulting inconsistency is expected to be numerically small, but to our knowledge its magnitude has not been measured.\linelabel{mismatch_sm:end}

\bibliography{references,footnotes}